\documentclass[twocolumn]{autart}     

\usepackage[dvipdfmx]{graphicx}
\usepackage{amssymb,mathrsfs,color}
\usepackage[cmex10]{amsmath}
\usepackage{array}
\usepackage{latexsym}
\usepackage{boxedminipage}
\usepackage{fancybox}
\usepackage{ascmac}
\usepackage{algorithm}
\usepackage{algorithmic}
\usepackage{amsmath}
\usepackage{cite}

\usepackage{boxedminipage}
\usepackage{fancybox}
\usepackage{ascmac}
\usepackage{url}

\usepackage[sort]{natbib}

\newtheorem{theorem}         {Theorem}[section]
\newtheorem{problem}             {Problem}[section]
\newtheorem{assumption}             {Assumption}[section]
\newtheorem{lemma}           {Lemma}[section]

\begin{document}

\setlength{\abovedisplayskip}{7pt}  
\setlength{\belowdisplayskip}{7pt}  

\begin{frontmatter}

\title{
Retrofit Control: Localization of Controller Design and Implementation 
\vspace{-10mm}
\thanksref{footnoteinfo}}  

\thanks[footnoteinfo]{
This work was supported by JST CREST Grant Number JP-MJCR15K1, Japan.
Corresponding author: T.~Ishizaki, Tel. \&\ Fax: +81-3-5734-2646. 
}

\author[TIT]{Takayuki Ishizaki$^{*}$}\ead{ishizaki@sc.e.titech.ac.jp},    
\author[TIT]{Tomonori Sadamoto}\ead{sadamoto@cyb.sc.e.titech.ac.jp},
\author[TIT]{Jun-ichi Imura}\ead{imura@sc.e.titech.ac.jp},
\author[KTH]{Henrik~Sandberg}\ead{hsan@kth.se},
and
\author[KTH]{Karl~Henrik~Johansson}\ead{kallej@kth.se},   

\address[TIT]{Tokyo Institute of Technology; 2-12-1, Ookayama, Meguro, Tokyo, 152-8552, Japan.}   
\address[KTH]{Department of Automatic Control, KTH Royal Institute of Technology, SE-100 44 Stockholm, Sweden.\vspace{-6mm}}



\begin{abstract}                           
In this paper, we propose a retrofit control method for stable network systems.
The proposed approach is a control method that, rather than an entire system model, requires a model of the subsystem of interest for controller design.
To design the retrofit controller, we use a novel approach based on hierarchical state-space expansion that generates a higher-dimensional cascade realization of a given network system.
The upstream dynamics of the cascade realization corresponds to an isolated model of the subsystem of interest, which is stabilized by a local controller.
The downstream dynamics can be seen as a dynamical model representing the propagation of interference signals among subsystems, the stability of which is equivalent to that of the original system.
This cascade structure enables a systematic analysis of both the stability and control performance of the resultant closed-loop system.
The resultant retrofit controller is formed as a cascade interconnection of the local controller and an output rectifier that rectifies an output signal of the subsystem of interest so as to conform to an output signal of the isolated subsystem model while acquiring complementary signals neglected in the local controller design, such as interconnection signals from neighboring subsystems.
Finally, the efficiency of the retrofit control method is demonstrated through numerical examples of power systems control and vehicle platoon control.
\vspace{-4mm}
\end{abstract}

\end{frontmatter}


\section{Introduction}

Recent developments in computer networking technology have enabled large-scale systems to be operated in a spatially distributed fashion.
For example, in power systems control \citep{kundur1994power}, a system operator manages distributed power plants with distributed  measurement units to meet the demands of a number of consumers. 
Towards the systematic control of such large-scale network systems, decentralized and distributed control techniques have been studied over the past half century; see \citep{siljak1991decentralized,vsiljak2005control} and the references therein.
In this line of study, there are found several illustrative results that highlight the difficulty of  controller design problems with structural constraints \citep{blondel2000survey,rotkowitz2006characterization}.

Starting from different perspectives, a number of decentralized and distributed control methods have been devised to overcome the difficulty of structured controller design. 
In this paper, we refer to structured control in which the subcontrollers have no direct communication among them as \textit{decentralized control} and structured control in which subcontrollers have communication with neighboring subcontrollers as \textit{distributed control}.
For example, \citep{siljak1972stability,wang1973stabilization,tan1990decentralized} report decentralized control methods on the basis of connective stability or related coprime factorization.
Furthermore, \citep{wang1995robust} introduces a decentralized control method based on small gain-type stability conditions or dissipation inequalities considering model uncertainty.
Similar dissipativity-based approaches are used in \citep{bamieh2002distributed,d2003distributed,langbort2004distributed} also for distributed control, and \citep{rantzer2015scalable} introduces a distributed control method for positive systems that has good scalability.
However, most existing decentralized and distributed control methods do not meet practical requirements, because they require an entire system model for controller design, and handle the design of all subcontrollers simultaneously.
In fact, for large-scale systems control, it is not generally reasonable to assume the availability of an entire system model, because subsystem parameters and controller structures may not be fully known  in the event of degradation, modification, and development of the subcontrollers and subsystems.
From this viewpoint, such \textit{centralized design} of decentralized and distributed controllers is impractical for large-scale systems, even though the resulting controller may be implemented in a distributed fashion.

To overcome this issue, the concept of \textit{distributed design} has been introduced in \citep{langbort2010distributed}, where the authors discuss the performance limitations of linear quadratic regulators designed in a distributed manner.
This result has been generalized to the case of networks composed of multi-dimensional subsystems, the states of which are fully controlled \citep{farokhi2013optimal}.
Furthermore, in \citep{ebihara2012decentralized}, a distributed design method for decentralized control using the $\mathcal{L}_{1}$-norm has been developed for positive linear systems.
Because each focuses on a particular class of systems, it is not simple to generalize their results to a broader class of systems.
As a related work, \citep{farokhi2015optimal} discusses the distributed design of optimal state-feedback controllers for discrete-time linear systems with stochastically-varying model parameters.
Even though the design of each subsystem controller is performed based on its local model information, the resultant optimal controller is a centralized controller in the sense that each subcontroller requires the feedback of full state information.

Another approach towards distributed design is control synthesis based on passivity, or, more generally, dissipativity and passivity shortage \citep{willems1972dissipativeI,willems1972dissipativeII,sepulchre2012constructive}.
It is known that appropriate interconnections of passive subsystems retain the passivity.
This implies that the entire network system can be guaranteed to be stable provided that each subsystem is individually designed to be passive.
However, in general, the design of subsystem interconnection structures is difficult to perform in a distributed manner.
For example, the interconnection matrix for passive subsystems is required to be negative semidefinite \citep{moylan1978stability}, and 
that for passivity-short subsystems is required to have a low-gain property in terms of eigenvalues in addition to negative semidefiniteness \citep{qu2014modularized}.
These characteristics are not fully determined by local interconnection structures.

With this background, the present paper develops a distributed design method for decentralized control that does not require an entire system model.
Instead, only a model of the subsystem of interest is needed for controller design, an approach that we call \textit{retrofit control}.
This retrofit control is based on the premise that a given network system, which can involve nonlinearity, is originally stable, and the interconnection signal flowing into the subsystem of interest is measurable.
It is shown that the resultant closed-loop system remains stable and its control performance can be improved with respect to a suitable measure.
This enables the scalable development of large-scale network systems because, towards further performance improvement, it is possible to consider the retrofit control of other subsystems while keeping the entire system stable.

To develop such a retrofit control method, we use a novel approach based on \textit{hierarchical state-space expansion}, which generates a higher-dimensional cascade realization of the given network system, called a \textit{hierarchical realization}.
Its upstream dynamics corresponds to an isolated model of the subsystem of interest, decoupled from the other subsystems.
A controller that stabilizes the isolated subsystem model is called a local controller.
The downstream dynamics can be seen as a dynamical model that represents the propagation of interference signals among subsystems, the stability of which is equivalent to that of the original network system.
It is shown that stabilization and improved control performance can be systematically realized.
The resultant retrofit controller, which measures a local output signal and an interconnection signal from neighboring subsystems, is formed as a cascade interconnection of the local controller designed for the isolated subsystem model and a dynamical rectifier, which we call an \textit{output rectifier}.
As a generalization of this result, we further consider removing the assumption of the interconnection signal measurements.
The resultant retrofit controller, which only measures the state of the subsystem of interest, also offers guaranteed stability and improved control performance.

The foundations of our contribution can be found in various previous studies.
Based on the inclusion principle, relevant to state-space expansion, a distributed control method has been developed in \citep{ikeda1984inclusion,iftar1993decentralized}.
Although some applications to vehicle control are described in \citep{stipanovic2004decentralized}, this method does not necessarily produce a stabilizing controller for general systems.
This limitation comes from the fact that a decentralized control design with an algebraic constraint is needed for an expanded system.
Moreover, the controller is designed in a centralized fashion.
This contrasts with the proposed retrofit control, which enables the systematic distributed design of decentralized control.
This paper builds on preliminary versions, unifying the results of hierarchical distributed control in \citep{sadamoto2014hierarchical} and nonlinear retrofit control \citep{sadamoto2016retrofitting} on the basis of the parameterized hierarchical state-space expansion.
This paper also provides detailed mathematical proofs and extensive numerical examples to underline the significance of the retrofit control.

Finally, we make a comparison with robust control \citep{zhou1996robust}.
In fact, localized controller design may be performed by a standard robust control method if all of the neighboring subsystems other than the subsystem of interest are regarded as model uncertainty.
However, this approach generally results in conservative consequences due to, e.g., the overestimation of uncertain system gains especially when available information on neighboring subsystems is limited.
In contrast, the retrofit control is just reliant on the stability of a given network system.
The retrofit controller guarantees robust stability in the sense that the entire closed-loop system is stable for any variations of neighboring subsystems other than the subsystem of interest, the norm bound of which is not assumed, as long as the given network system is originally stable.

The remainder of this paper is organized as follows.
In Section~\ref{secpf1}, we formulate a fundamental problem of retrofit control.
Then, in Section~\ref{secrcsls}, hierarchical state-space expansion is introduced to solve it.
Section~\ref{secrem} discusses the generalization of the proposed approach to nonlinear systems, amongst other remarks.
In Section~\ref{secpf2}, we formulate a retrofit control problem without the assumption of  interconnection signal measurements, and then we provide a solution in Section~\ref{secsol2}.
Section~\ref{secnumex} contains numerical examples of power systems and vehicle platoon control, demonstrating the results in Sections~\ref{secfrc} and \ref{secgrc}, respectively.
Finally, concluding remarks are given in Section~\ref{seccr}.

\textbf{Notation}~
We denote the set of real numbers by $\mathbb{R}$, the identity matrix by $I$, the transpose of a matrix $M$ by $M^{{\sf T}}$, the image of a matrix $M$ by ${\rm im}\ \!M$, the kernel by ${\rm ker}\ \!M$, a left inverse of a left invertible matrix $P$ by $P^{\dagger}$, the $\mathcal{L}_{2}$-norm of a square-integrable function $f$ by $\|f\|_{\mathcal{L}_{2}}$,
the $\mathcal{H}_{2}$-norm of a stable proper transfer matrix $G$ by $\|G\|_{\mathcal{H}_{2}}$, and the $\mathcal{H}_{\infty}$-norm of a stable transfer matrix $G$ by $\|G\|_{\mathcal{H}_{\infty}}$.
A map $\mathcal{F}$ is said to be a dynamical map if the triplet $(x,u,y)$ with $y=\mathcal{F}(u)$ solves a system of differential equations 
\[
\dot{x}=f(x,u)\quad y=g(x,u) 
\]
with some functions $f$ and $g$, and an initial value $x(0)$.


\section{Fundamentals of Retrofit Control}\label{secfrc}

\subsection{Problem Formulation}\label{secpf1}

Consider an interconnected linear system described by
\begin{subequations}\label{sys12}
\begin{align}
&\Sigma_{1}:\left\{
\begin{array}{ccl}
\dot{x}_{1}&\hspace{-0pt}=&\hspace{-0pt}\mbox{\boldmath $A$}_{1}x_{1}+
\mbox{\boldmath $L$}_{1}\gamma_{2}+\mbox{\boldmath $B$}_{1}u_{1}\vspace{-1mm}\\
y_{1}&\hspace{-0pt}=&\hspace{-0pt}\mbox{\boldmath $C$}_{1}x_{1}
\end{array}
\right.\label{sys1}\vspace{-2mm}\\
&\Sigma_{2}:\left\{
\begin{array}{ccl}
\dot{x}_{2}&\hspace{-0pt}=&\hspace{-0pt}A_{2}x_{2}+L_{2}{\mathit\Gamma}_{1}x_{1}\vspace{-1mm}\\
\gamma_{2}&\hspace{-0pt}=&\hspace{-0pt}{\mathit\Gamma}_{2}x_{2}
\end{array}
\right.\label{sys2}
\end{align}
\end{subequations}
where $x_{1}$ and $x_{2}$ denote the states of $\Sigma_{1}$ and $\Sigma_{2},\ u_{1}$ and $y_{1}$ denote the external input signal and the measurement output signal of $\Sigma_{1}$, and $\gamma_{2}$ denotes the interconnection signal of $\Sigma_{2}$ injected into $\Sigma_{1}$.
The dimensions of $\Sigma_{1}$ and $\Sigma_{2}$ are denoted by $n_{1}$ and $n_{2}$, respectively.

In the following, based on the premise that the system model of $\Sigma_{1}$ is available but that of $\Sigma_{2}$ is not, we consider the design of a controller implemented to $\Sigma_{1}$.
We refer to such a controller as a \textit{retrofit controller}, whereby the design and implementation are both localized with the subsystem of interest, i.e., $\Sigma_{1}$.
Throughout this paper, the system parameters available for retrofit controller design are represented by symbols in bold face, such as $\mbox{\boldmath $A$}_{1}$,  $\mbox{\boldmath $B$}_{1}$,  $\mbox{\boldmath $C$}_{1}$, and $\mbox{\boldmath $L$}_{1}$ in (\ref{sys1}).
As seen in Section~\ref{secgnc}, $\Sigma_{2}$ can be generalized to a nonlinear system.

Describing the interconnected system of (\ref{sys1}) and (\ref{sys2}) as
\begin{equation}\label{prsys}
\Sigma:
\left\{
\begin{array}{ccl}
\left[\hspace{0pt}
\begin{array}{cc}
\dot{x}_{1}\\
\dot{x}_{2}
\end{array}
\hspace{0pt}
\right]
&\hspace{0pt}=&\hspace{0pt}
\left[\hspace{0pt}
\begin{array}{cc}
\mbox{\boldmath $A$}_{1}&\hspace{0pt}\mbox{\boldmath $L$}_{1}{\mathit \Gamma}_{2}\\
 L_{2}{\mathit \Gamma}_{1}&\hspace{0pt}A_{2}
\end{array}
\hspace{0pt}
\right]
\left[\hspace{0pt}
\begin{array}{cc}
x_{1}\\
x_{2}
\end{array}
\hspace{0pt}
\right]
+
\left[\hspace{0pt}
\begin{array}{cc}
\mbox{\boldmath $B$}_{1}\\
0
\end{array}
\hspace{0pt}
\right]
u_{1}\vspace{0pt}\\
y_{1}&\hspace{0pt}=&\hspace{0pt}
\left[\hspace{0pt}
\begin{array}{cc}
\mbox{\boldmath $C$}_{1}&\hspace{0pt}0
\end{array}
\hspace{0pt}
\right]
\left[\hspace{0pt}
\begin{array}{cc}
x_{1}\\
x_{2}
\end{array}
\hspace{0pt}
\right],
\end{array}
\right. 
\end{equation}
we refer to (\ref{prsys}) as the \textit{preexisting system}.
To clarify the subsequent discussion, the assumptions for the retrofit controller design can be stated as follows:

\begin{assumption}\label{assump}\normalfont
For the preexisting system $\Sigma$ in (\ref{prsys}), the following assumptions are made.
\begin{description}
\item[(i)] The preexisting system $\Sigma$ is internally stable, i.e.,
\begin{equation}\label{asclsta}
A:=\left[\hspace{-0pt}
\begin{array}{cc}
\mbox{\boldmath $A$}_{1}&\hspace{-0pt}\mbox{\boldmath $L$}_{1}{\mathit \Gamma}_{2}\\
L_{2}{\mathit \Gamma}_{1}&\hspace{-0pt}A_{2}
\end{array}
\hspace{-0pt}
\right]
\end{equation}
is stable.
\item[(ii)] For the design of a retrofit controller, the system matrices of $\Sigma_{1}$, i.e., the bold face matrices in (\ref{sys1}), are available, but those of $\Sigma_{2}$ in (\ref{sys2}) are not.
\item[(iii)] For the implementation of a retrofit controller, the measurement output signal $y_{1}$ and the interconnection signal $\gamma_{2}$ are measurable.
\end{description}
\end{assumption}

Assumption~\ref{assump}~\textbf{(i)} implies that the internal stability of the preexisting system has been assured before implementing a retrofit controller.
This assumption is reasonable when we consider retrofit control for a stably operated system, where 
 a preexisting stabilizing controller can be involved in $\Sigma_{2}$.
Assumption~\ref{assump}~\textbf{(ii)} is concerned with the localization ability of controller design.
This assumption implies that we are only allowed to use the local information of the system model of $\Sigma_{1}$ for the retrofit controller design.
Assumption~\ref{assump}~\textbf{(iii)} is concerned with the localization ability of controller implementation, which is usually discussed in the context of distributed control for reducing the communication and computation costs of controller implementation.

The objective of the proposed retrofit control method is to improve control performance with respect to a suitable measure.
To simplify the discussion, let us consider a situation where an unknown state deflection arises in $\Sigma_{1}$ at some instant.
This can be described as a transient system response with the initial condition
\begin{equation}\label{stdef}
x_{1}(0)=\delta_{0},\quad x_{2}(0)=0
\end{equation}
where $\delta_{0}$ corresponds to the state deflection.
Without loss of generality, we assume that $\delta_{0}$ is contained in the unit ball denoted by
\[
\mathcal{B}=\{\delta_{0}\in \mathbb{R}^{n_{1}}:\|\delta_{0}\|\leq 1\}.
\]
Note that disturbance attenuation with an evaluation output can be addressed in a similar manner by setting a disturbance input port on $\Sigma_{1}$.
In this formulation, we address the following retrofit controller design problem.

\begin{problem}\label{prob1}\normalfont
Consider the preexisting system $\Sigma$ in (\ref{prsys}) with the initial condition (\ref{stdef}).
Under Assumption~\ref{assump}, find a retrofit controller of the form
\begin{equation}\label{retform}
\Pi_{1}:u_{1}=\mathcal{K}_{1}(y_{1},\gamma_{2}),
\end{equation}
where $\mathcal{K}_{1}$ denotes a dynamical map, such that 
\begin{description}
\item[(A)] the closed-loop system composed of (\ref{prsys}) and (\ref{retform}) is internally stable for any $\Sigma_{2}$ such that $\Sigma$ is internally stable, and
\item[(B)] for any state deflection $\delta_{0}\in \mathcal{B}$, the magnitude of $\|x_{1}\|_{\mathcal{L}_{2}}$ and $\|x_{2}\|_{\mathcal{L}_{2}}$ is sufficiently small with respect to a suitable threshold.
\end{description}
\end{problem}

The initial condition (\ref{stdef}) represents a local disturbance injected into $\Sigma_{1}$ in (\ref{sys1}).
This can be regarded as an impulsive variation of the subsystem state, which can model, e.g., three-phase faults in power systems control \citep{kundur1994power}.
The objective of the retrofit controller $\Pi_{1}$ in (\ref{retform}) is to attenuate the impact of the local disturbance on the subsystem $\Sigma_{1}$ and limit the propagation to the other subsystem, i.e., $\Sigma_{2}$.

\begin{figure}[t]
\begin{center}
\includegraphics[width=50mm]{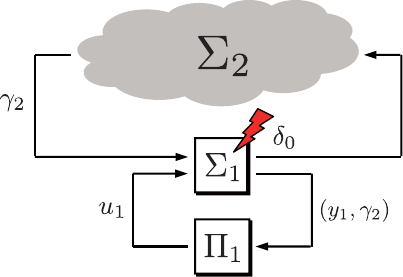}
\end{center}
\vspace{-4pt}
\caption{\scriptsize
Signal-flow diagram of retrofit control.}
\label{figsretrofit}
\vspace{-4pt}
\end{figure}

A schematic depiction of this retrofit control is shown in Fig.~\ref{figsretrofit}.
Note that $\Sigma_{2}$ in Fig.~\ref{figsretrofit} can itself be regarded as a large-scale network system composed of preexisting subcontrollers and subsystems, because its dimension and structure have no limitation in this formulation.
In general, it is not realistic to assume that an entire system model is available for large-scale network systems.
In addition, the simultaneous design of all subcontrollers is generally difficult for large-scale network systems control.
Even though $\Sigma_{2}$ may be regarded as model uncertainty, it is typically assumed to be norm-bounded in  robust control.
The retrofit control problem, seeking a controller that guarantees the closed-loop system stability for all possible $\Sigma_{2}$ such that the preexisting system $\Sigma$ is stable, is different from usual robust control problems \citep{zhou1996robust}.

As we have stated, the retrofit control method does not require an entire system model. 
Instead, we use only the system model of $\Sigma_{1}$ for controller design.
The resultant closed-loop system is required to be stable provided that the preexisting system is originally stable, and its control performance is to be improved.
For further performance improvements, one can consider applying retrofit control to other subsystems involved in $\Sigma_{2}$, while keeping the entire system stable, i.e., \textit{distributed design} of multiple retrofit controllers.
This enables the scalable development of large-scale network systems; see Section~\ref{secsmrc} for further details.

\subsection{Solution via Hierarchical State-Space Expansion}\label{secrcsls}

Towards the systematic design of a retrofit controller, we introduce a state-space expansion technique, called \textit{hierarchical state-space expansion}.

\begin{lemma}\label{lemhr}\normalfont
For the preexisting system $\Sigma$ in (\ref{prsys}), consider the cascade interconnection system whose upstream subsystem is given by
\begin{subequations}\label{hrex}
\begin{equation}\label{hrup}
\dot{\hat{\xi}}_{1}=\mbox{\boldmath $A$}_{1}\hat{\xi}_{1}+\mbox{\boldmath $B$}_{1}u_{1},
\end{equation}
which is $n_{1}$-dimensional, and downstream subsystem is given by
\begin{equation}\label{hrdn}
\left[\hspace{0pt}
\begin{array}{cc}
\dot{\xi}_{1}\\
\dot{\xi}_{2}
\end{array}
\hspace{0pt}
\right]
=
\left[\hspace{0pt}
\begin{array}{cc}
\mbox{\boldmath $A$}_{1}&\hspace{0pt}\mbox{\boldmath $L$}_{1}{\mathit \Gamma}_{2}\\
 L_{2}{\mathit \Gamma}_{1}&\hspace{0pt}A_{2}
\end{array}
\hspace{0pt}
\right]
\left[\hspace{0pt}
\begin{array}{cc}
\xi_{1}\\
\xi_{2}
\end{array}
\hspace{0pt}
\right]
+
\left[\hspace{0pt}
\begin{array}{cc}
0\\
 L_{2}{\mathit \Gamma}_{1}
\end{array}
\hspace{0pt}
\right]\hat{\xi}_{1},
\end{equation}
\end{subequations}
which is $(n_{1}+n_{2})$-dimensional.
Then,
\begin{equation}\label{stateeq}
x_{1}(t)=\xi_{1}(t)+\hat{\xi}_{1}(t),\quad x_{2}(t)=\xi_{2}(t),\quad\forall t\geq 0
\end{equation}
for any external input signal $u_{1}$, provided that (\ref{stateeq}) is satisfied at the initial time $t=0$.
\end{lemma}

We can easily verify the claim by summing the differential equations (\ref{hrup}) and (\ref{hrdn}).
Hierarchical state-space expansion in Lemma~\ref{lemhr} produces a higher-dimensional cascade realization composed of the upstream dynamics (\ref{hrup}) and the downstream dynamics (\ref{hrdn}), which is a $(2n_{1}+n_{2})$-dimensional system.
We refer to  (\ref{hrex}) as a \textit{hierarchical realization} of the preexisting system $\Sigma$.
Note that the upstream dynamics (\ref{hrup}) can be regarded as the isolated model of $\Sigma_{1}$, whose system matrices are assumed to be available; see Assumption~\ref{assump}~\textbf{(ii)}.
In contrast, the downstream dynamics (\ref{hrdn}) can be seen as a dynamical model representing the propagation of the interconnection signal from $\Sigma_{1}$.
Note that the downstream dynamics (\ref{hrdn}) is internally stable because, the preexisting system $\Sigma$ is assumed to be internally stable; see Assumption~\ref{assump}~\textbf{(i)}.

For consistency with (\ref{stdef}) and (\ref{stateeq}), we describe the initial condition of the hierarchical realization (\ref{hrex}) as
\begin{equation}\label{inicon}
\hat{\xi}_{1}(0)=\delta_{0}-\zeta_{0},\quad\left[\hspace{0pt}
\begin{array}{cc}
\xi_{1}(0)\\
\xi_{2}(0)
\end{array}
\hspace{0pt}
\right]=
\left[\hspace{0pt}
\begin{array}{cc}
\zeta_{0}\\
0
\end{array}
\hspace{0pt}
\right],
\end{equation}
where $\zeta_{0}\in \mathbb{R}^{n_{1}}$ can be seen as an arbitrary parameter.
On the basis of this, we consider the design of a local controller for the upstream dynamics (\ref{hrup}), namely, the isolated model of the subsystem of interest.
For simplicity, we assume that the local controller is designed as a static output feedback controller
\begin{equation}\label{stacon}
u_{1}=\mbox{\boldmath $K$}_{1}\mbox{\boldmath $C$}_{1}\hat{\xi}_{1}.
\end{equation}
More specifically, this local controller is designed such that the closed-loop dynamics
\begin{equation}\label{xihsta}
\dot{\hat{\xi}}_{1}=(\mbox{\boldmath $A$}_{1}+\mbox{\boldmath $B$}_{1}\mbox{\boldmath $K$}_{1}\mbox{\boldmath $C$}_{1})\hat{\xi}_{1}
\end{equation}
is internally stable and the control performance specification
\begin{equation}\label{xihprf}
\|\hat{\xi}_{1}\|_{\mathcal{L}_{2}}\leq\epsilon_{1},\quad\forall\hat{\xi}_{1}(0)\in \mathcal{B}
\end{equation}
is satisfied for a given tolerance $\epsilon_{1}>0$.
In fact, generalization to the design of dynamical output feedback controllers is straightforward; see Section~\ref{secsub2}.

Based on the cascade structure of (\ref{hrex}), the stability and control performance of the closed-loop system can be easily analyzed as follows.

\begin{lemma}\label{lemspec}\normalfont
For the hierarchical realization (\ref{hrex}), consider the local output feedback controller (\ref{stacon}).
Under Assumption~\ref{assump}~\textbf{(i)}, the closed-loop system composed of (\ref{hrex}) and (\ref{stacon}) is internally stable if and only if the closed-loop dynamics (\ref{xihsta}) is internally stable.
Furthermore, for $i\in\{1,2\}$, let
\begin{equation}\label{Gis}
G_{i}(s):=E_{i}^{{\sf T}}(sI-A)^{-1}E_{2}L_{2}{\mathit \Gamma}_{1}
\end{equation}
denote the transfer matrix from $\hat{\xi}_{1}$ to $\xi_{i}$ of the downstream dynamics (\ref{hrdn}), where $A$ is defined as in (\ref{asclsta}), and
\begin{equation}\label{defE}
E_{1}:=\left[\hspace{0pt}
\begin{array}{cc}
I\\
0
\end{array}
\hspace{0pt}
\right],\quad
E_{2}:=\left[\hspace{0pt}
\begin{array}{cc}
0\\
I
\end{array}
\hspace{0pt}
\right].
\end{equation}
If (\ref{xihprf}) holds for the closed-loop dynamics (\ref{xihsta}), then
\begin{equation}\label{prfbnd}
\hspace{0pt}
\begin{array}{rcl}
\|\xi_{1}+\hat{\xi}_{1}\|_{\mathcal{L}_{2}}&\hspace{0pt}\leq&\hspace{0pt}
\alpha_{1}(1+\|\zeta_{0}\|)\epsilon_{1}+\beta_{1}(\zeta_{0}),\vspace{-0mm}\\
\|\xi_{2}\|_{\mathcal{L}_{2}}&\hspace{0pt}\leq&\hspace{0pt}\alpha_{2}(1+\|\zeta_{0}\|)\epsilon_{1}+\beta_{2}(\zeta_{0}),
\end{array}
\quad\forall\delta_{0}\in \mathcal{B}\hspace{-10pt}
\end{equation}
with the initial condition (\ref{inicon}), where the nonnegative constants
\begin{equation}\label{mus}
\alpha_{1}:=\|G_{1}+I\|_{\mathcal{H}_{\infty}},\quad\alpha_{2}:=\|G_{2}\|_{\mathcal{H}_{\infty}}
\end{equation}
and the nonnegative functions
\begin{equation}\label{betai}
\beta_{i}(\zeta_{0}):=\|E_{i}^{{\sf T}}e^{At}E_{1}\zeta_{0}\|_{\mathcal{L}_{2}}
\end{equation}
are independent of the selection of the feedback gain $\mbox{\boldmath $K$}_{1}$ in (\ref{stacon}).
\end{lemma}\vspace{-2mm}

\begin{pf}
Owing to the cascade structure of the hierarchical realization, the internal stability of the closed-loop system composed of (\ref{hrex}) and (\ref{stacon}) is equivalent to that of (\ref{xihsta}), provided that Assumption~\ref{assump}~\textbf{(i)} holds.
Furthermore, let
\[
X_{1}(s):=\bigl(sI-(\mbox{\boldmath $A$}_{1}+
\mbox{\boldmath $B$}_{1}\mbox{\boldmath $K$}_{1}\mbox{\boldmath $C$}_{1})\bigr)^{-1}(\delta_{0}-\zeta_{0})
\]
denote the Laplace transform of $\hat{\xi}_{1}$ in (\ref{xihsta}) with the initial condition (\ref{inicon}).
Note that
$\|X_{1}\|_{\mathcal{H}_{2}}\leq\bigl(1+\|\zeta_{0}\|\bigr)\epsilon_{1}$ for all $\delta_{0}\in \mathcal{B}$ if (\ref{xihprf}) holds.
Then, we see that
$(G_{1}+I)X_{1}$ corresponds to the Laplace transforms of $\xi_{1}+\hat{\xi}_{1}$ and $G_{2}X_{1}$ corresponds to that of $\xi_{2}$ when we restrict the initial condition to
\[
\hat{\xi}_{1}(0)=\sigma(\delta_{0}-\zeta_{0}),\quad
\left[\hspace{0pt}
\begin{array}{cc}
\xi_{1}(0)\\
\xi_{2}(0)
\end{array}
\hspace{0pt}
\right]=(1-\sigma)
\left[\hspace{0pt}
\begin{array}{cc}
\zeta_{0}\\
0
\end{array}
\hspace{0pt}
\right]
\]
with $\sigma=1$.
In addition, when we restrict the initial condition to the case of $\sigma=0$, 
the time evolution of the downstream dynamics (\ref{hrdn}) given as $e^{At}E_{1}\zeta_{0}$ is independent of the closed-loop dynamics (\ref{xihsta}).
Thus, (\ref{prfbnd}) follows from the cascade structure of (\ref{hrex}). \hfill$\square $
\end{pf}\vspace{-2mm}

As stated in Lemma~\ref{lemspec}, the nonnegative constants $\alpha_{i}$ and functions $\beta_{i}$, which are relevant to the system matrices of the preexisting system $\Sigma$ in (\ref{prsys}) and the parameter $\zeta_{0}$ in (\ref{inicon}), are independent of the local controller design of (\ref{stacon}).
Thus, in designing a local controller such that the bound (\ref{xihprf}) is satisfied for a smaller tolerance $\epsilon_{1}$, we can attain improved control performance in the sense of the upper bounds in (\ref{prfbnd}).
Note that (\ref{prfbnd}) implies the bounds of $\|x_{1}\|_{\mathcal{L}_{2}}$ and $\|x_{2}\|_{\mathcal{L}_{2}}$ owing to the relation of (\ref{stateeq}).
Clearly, the minimum values of the bounds are given by $\alpha_{i}\epsilon_{1}$ when we take $\zeta_{0}$ in (\ref{inicon}) as
\begin{equation}\label{zeta0}
\zeta_{0}=0.
\end{equation}
Thus, in the following, we focus our attention on the initial condition (\ref{inicon}) with this selection of $\zeta_{0}$.

It remains to demonstrate the implementation of the local output feedback controller (\ref{stacon}) for the original realization $\Sigma$ in (\ref{prsys}).
Note that the output signal $\mbox{\boldmath $C$}_{1}\hat{\xi}_{1}$ from the hierarchical realization is not directly measurable from the original realization.
To generate $\mbox{\boldmath $C$}_{1}\hat{\xi}_{1}$ for controller implementation, we introduce a dynamical memory, which we call an \textit{output rectifier}, that achieves
\begin{equation}\label{xhat1}
\hat{x}_{1}(t)=\xi_{1}(t),\quad\forall t\geq 0,
\end{equation}
where $\hat{x}_{1}$ denotes the state of the output rectifier.
Based on the fact that $\gamma_{2}={\mathit \Gamma}_{2}\xi_{2}$ in the dynamics of $\xi_{1}$ of (\ref{hrdn}), such an output rectifier can be realized as
\begin{equation}\label{dyncomp}
\left\{
\begin{array}{ccl}
\dot{\hat{x}}_{1}&\hspace{0pt}=&\hspace{0pt}\mbox{\boldmath $A$}_{1}\hat{x}_{1}+\mbox{\boldmath $L$}_{1}\gamma_{2}\vspace{-1mm}\\
\hat{y}_{1}&\hspace{0pt}=&\hspace{0pt}y_{1}-\mbox{\boldmath $C$}_{1}\hat{x}_{1},
\end{array}
\right.
\end{equation}
whose initial condition is determined by (\ref{zeta0}) as
\begin{equation}\label{inixh}
\hat{x}_{1}(0)=0.
\end{equation}
This initial condition is actually consistent with (\ref{inicon}) and (\ref{xhat1}).
In fact, with this $n_{1}$-dimensional output rectifier, the output signal $\mbox{\boldmath $C$}_{1}\hat{\xi}_{1}$ can be generated as $\hat{y}_{1}$ in (\ref{dyncomp}) based on the relation on the left of (\ref{stateeq}). 
In conclusion, a solution to Problem~\ref{prob1} is given as follows.

\begin{theorem}\label{thm1}\normalfont
Under Assumption~\ref{assump}~\textbf{(i)}, consider the preexisting system $\Sigma$ in (\ref{prsys}) with the initial condition (\ref{stdef}).
For any local output feedback controller in (\ref{stacon}) such that the closed-loop dynamics (\ref{xihsta}) is internally stable and (\ref{xihprf}) holds, the entire closed-loop system composed of (\ref{prsys}) and
\begin{equation}\label{rcons}
\Pi_{1}:\left\{
\begin{array}{ccl}
\dot{\hat{x}}_{1}&\hspace{0pt}=&\hspace{0pt}\mbox{\boldmath $A$}_{1}\hat{x}_{1}+\mbox{\boldmath $L$}_{1}\gamma_{2}\vspace{-1mm}\\
u_{1}&\hspace{0pt}=&\hspace{0pt}\mbox{\boldmath $K$}_{1}(y_{1}-\mbox{\boldmath $C$}_{1}\hat{x}_{1})
\end{array}
\right.
\end{equation}
with the initial condition (\ref{inixh}) is internally stable and
\begin{equation}\label{prfbndx}
\|x_{1}\|_{\mathcal{L}_{2}}\leq\alpha_{1}\epsilon_{1},\quad\|x_{2}\|_{\mathcal{L}_{2}}\leq\alpha_{2}\epsilon_{1},\quad\forall\delta_{0}\in \mathcal{B},
\end{equation}
where $\alpha_{1}$ and $\alpha_{2}$ in (\ref{mus}) are independent of the local controller design of (\ref{stacon}).
\end{theorem}\vspace{-2mm}

\begin{pf}
As stated in Lemma~\ref{lemspec}, the closed-loop system in the hierarchical realization, i.e., (\ref{hrex}) with (\ref{stacon}), is internally stable.
Note that the closed-loop system in the original realization, i.e., (\ref{prsys}) with (\ref{rcons}), is related to the closed-loop system in the hierarchical realization by (\ref{stateeq}) and (\ref{xhat1}).
This can be regarded as the coordinate transformation, i.e., the bijection, from the hierarchical realization to the original realization.
The inverse of this transformation is given by
\begin{equation}\label{ctrans}
\left[\hspace{0pt}
\begin{array}{cc}
\xi_{1}\\
\xi_{2}\\
\hat{\xi}_{1}
\end{array}
\hspace{0pt}
\right]
=
\left[\hspace{0pt}
\begin{array}{ccc}
0&0&I\\
0&I&0\\
I&0&-I
\end{array}
\hspace{0pt}
\right]
\left[\hspace{0pt}
\begin{array}{cc}
x_{1}\\
x_{2}\\
\hat{x}_{1}
\end{array}
\hspace{0pt}
\right].
\end{equation}
Thus, their internal stability is equivalent.
This also shows that (\ref{prfbnd}) with (\ref{zeta0}) is equivalent to (\ref{prfbndx}).\hfill$\square $
\end{pf}\vspace{-2mm}

Theorem~\ref{thm1} shows that the $\mathcal{L}_{2}$-norm of the transient state response is improved in the sense of the upper bound in (\ref{prfbndx}) by designing a local controller such that (\ref{xihprf}) is satisfied for a smaller tolerance $\epsilon_{1}$, even though the exact values of $\alpha_{1}$ and $\alpha_{2}$ are not available because the system model of $\Sigma_{2}$ is assumed to be unavailable.
The resultant retrofit controller $\Pi_{1}$ in (\ref{rcons}) is formed as the cascade interconnection of the local output feedback controller (\ref{stacon}) and the $n_{1}$-dimensional output rectifier (\ref{dyncomp}).
The design and implementation of the retrofit controller comply with Assumptions~\ref{assump}~\textbf{(ii)} and \textbf{(iii)}.

A decentralized controller can be made as $u_{1}=\mbox{\boldmath $K$}_{1}y_{1}$, where $\mbox{\boldmath $K$}$ is designed based on the system model of $\Sigma_{1}$ as in (\ref{xihsta}).
However, this does not generally ensure the stability of the resultant closed-loop system, even if $\mbox{\boldmath $K$}_{1}$ is designed such that (\ref{xihsta}) is stable.
This is because the interconnection signal $\gamma_{2}$, neglected in the local controller design, affects the measurement output signal $y_{1}$ of $\Sigma_{1}$ and may induce undesirable output feedback.
To avoid such feedback, the output rectifier provides the compensation signal $\mbox{\boldmath $C$}_{1}\hat{x}_{1}$ to the local controller while measuring the interconnection signal $\gamma_{2}$.
The output rectifier can be regarded as a dynamical simulator to cancel out the interference of $\Sigma_{2}$ with the output signal $y_{1}$, the function of which is different from that of usual state observers and estimators.

\begin{figure}[t]
\begin{center}
\includegraphics[width=80mm]{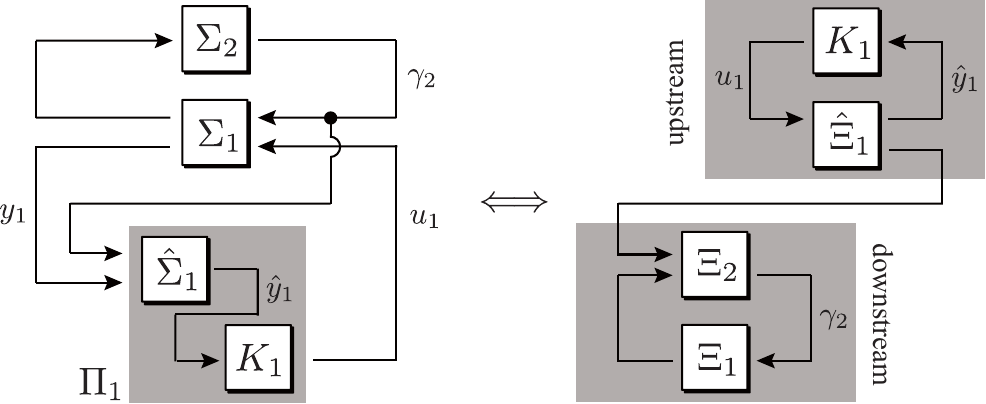}
\end{center}
\vspace{-0pt}
\caption{\scriptsize
Retrofit controller resulting from hierarchical realization.}
\label{figlogic}
\end{figure}

Fig.~\ref{figlogic} shows  schematic depiction of the signal flow diagram of the retrofit control and an equivalent diagram in the hierarchical realization. 
In the left diagram, the feedback loop of the blocks of $\Sigma_{1}$ and $\Sigma_{2}$ corresponds to the preexisting system (\ref{prsys}), and the shadowed block of $\Pi_{1}$ corresponds to the retrofit controller (\ref{rcons}).
The block of $\hat{\Sigma}_{1}$ represents the output rectifier (\ref{dyncomp}), and the block of $K_{1}$ represents the local controller $u_{1}=\mbox{\boldmath $K$}_{1}\hat{y}_{1}$.

In the right diagram, the feedback loop of the upper shadowed block corresponds to the closed-loop dynamics (\ref{xihsta}), where the upstream dynamics (\ref{hrup}) in the hierarchical realization is represented by the block of $\hat{\Xi}_{1}$ and the local output feedback controller (\ref{stacon}) is represented by the block of $K_{1}$.
The feedback loop of the lower shadowed block corresponds to the subsystems of the downstream dynamics (\ref{hrdn}), which are represented by the blocks of $\Xi_{1}$ and $\Xi_{2}$, respectively.
The equivalence between two diagrams is shown as the coordinate transformation in (\ref{ctrans}).

\subsection{Several Remarks}\label{secrem}

\subsubsection{Initial Condition Selection}\label{secsub1}

Owing to the internal stability of the closed-loop system shown in Theorem~\ref{thm1}, the selection of initial conditions for the output rectifier does not affect the stability of the closed-loop system.
In fact, for any initial conditions of $\Sigma_{1},\ \Sigma_{2}$ in (\ref{sys12}) and $\Pi_{1}$ in (\ref{rcons}), denoted by $x_{1}(0)$,  $x_{2}(0)$, and $\hat{x}_{1}(0)$, the initial condition of the hierarchical realization (\ref{hrex}) is uniquely determined as
being consistent with (\ref{stateeq}) and (\ref{xhat1}) or, equivalently, (\ref{ctrans}).
Note that $\hat{x}_{1}(0)=\zeta_{0}$,
which means that the free parameter $\zeta_{0}$ in (\ref{inicon}) corresponds to the initial condition of the output rectifier.
This shows the equivalence between (\ref{zeta0}) and (\ref{inixh}).

\subsubsection{Implementation of Multiple Retrofit Controllers}\label{secsmrc}
Under the output rectifier initial condition (\ref{inixh}), let us discuss the case where $x_{2}(0)$ is nonzero.
In particular, we first consider the case of $\delta_{0}=0$, which implies
\[
\hat{x}_{1}(0)=x_{1}(0)=0,
\]
i.e., the initial conditions of both subsystem $\Sigma_{1}$ and the output rectifier are zero.
In this situation, $\hat{x}_{1}(t)=x_{1}(t)$ or, equivalently, $\hat{\xi}_{1}(t)=0$ holds for all $t\geq 0$.
This is because the subsystem state $x_{1}$ and the output rectifier state $\hat{x}_{1}$ are equally driven by the interconnection signal $\gamma_{2}$ from $\Sigma_{2}$, whose initial condition is now assumed to be nonzero.
Therefore, the retrofit controller $\Pi_{1}$ does not take any control action, i.e., $u_{1}(t)=0$ for all $t\geq 0$, irrespective of the initial conditions of $\Sigma_{2}$.
Note that such state deflections of $\Sigma_{2}$ can be managed by another retrofit controller implemented in the corresponding subsystem.

Next, we consider the case where both $\delta_{0}$ and $x_{2}(0)$ are nonzero.
In a similar manner to that in Theorem~\ref{thm1}, we can derive the corresponding upper bound for the transient state response of $\Sigma_{1}$ as
\[
\|x_{1}\|_{\mathcal{L}_{2}}\leq\alpha_{1}\epsilon_{1}+\|E_{1}^{{\sf T}}e^{At}E_{2}x_{2}(0)\|_{\mathcal{L}_{2}},\quad\forall\delta_{0}\in \mathcal{B}.
\]
Note that the offset term relevant to $x_{2}(0)$ is not dependent on the selection of the feedback gain $\mbox{\boldmath $K$}_{1}$ in (\ref{stacon}).
When the subsystem $\Sigma_{2}$ is itself a network system composed of several subsystems, we can consider the simultaneous implementation of retrofit controllers to each of the respective subsystems.
This implies that multiple subsystem operators can independently plug in, plug out, and modify local controllers for the respective subsystems without concerning the instability of the entire network system.

\subsubsection{Local Dynamical Controller Design}\label{secsub2}

Next, let us consider the situation where a dynamical output feedback controller is designed, rather than the local static controller (\ref{stacon}).
This generalization can be done by simply replacing (\ref{stacon}) with
\begin{equation}\label{dycon}
u_{1}=\mathcal{K}_{1}(\mbox{\boldmath $C$}_{1}\hat{\xi}_{1})
\end{equation}
where $\mathcal{K}_{1}$ denotes the dynamical map of the local controller.
The controller design and implementation can only be performed by the system model of $\Sigma_{1}$.
Note that any conventional method can be applied for the design of a local dynamical controller (\ref{dycon}) that complies with the specification on internal stability in (\ref{xihsta}) and that on control performance in (\ref{xihprf}). 
The resultant retrofit controller is given by replacing $\mbox{\boldmath $K$}_{1}$ in (\ref{rcons}) with $\mathcal{K}_{1}$.
For example, if we design the dynamical map $\mathcal{K}_{1}$ in (\ref{dycon}) as an observer-based state feedback controller, then the retrofit controller is
\begin{equation}\label{obsretro}
\left\{
\begin{array}{ccl}
\dot{\hat{x}}_{1}&\hspace{0pt}=&\hspace{0pt}\mbox{\boldmath $A$}_{1}\hat{x}_{1}+\mbox{\boldmath $L$}_{1}\gamma_{2}\vspace{-1.5mm}\\
\dot{\zeta}_{1}&\hspace{0pt}=&\hspace{0pt}\mbox{\boldmath $A$}_{1}\zeta_{1}+\mbox{\boldmath $B$}_{1}u_{1}+
\mbox{\boldmath $H$}_{1}(y_{1}-\mbox{\boldmath $C$}_{1}\hat{x}_{1}-\mbox{\boldmath $C$}_{1}\zeta_{1})\vspace{-1.5mm}\\
u_{1}&\hspace{0pt}=&\hspace{0pt}\mbox{\boldmath $F$}_{1}\zeta_{1},
\end{array}
\right.\hspace{-10pt}
\end{equation}
where the feedback gains $\mbox{\boldmath $H$}_{1}$ and $\mbox{\boldmath $F$}_{1}$ are designed such that the specifications are satisfied for the isolated model of $\Sigma_{1}$.

\subsubsection{Generalization to Nonlinear Systems}\label{secgnc}

Because we do not use the system model of $\Sigma_{2}$ in (\ref{sys2}) for the retrofit controller design, we can generalize our approach to nonlinear systems.
More specifically, we consider replacing $\Sigma_{2}$ with
\begin{equation}\label{nlsys}
\dot{x}_{2}=f_{2}(x_{2},x_{1}),\quad
\gamma_{2}=h_{2}(x_{2},x_{1})
\end{equation}
where $f_{2}$ and $h_{2}$ denote some nonlinear functions.
The corresponding preexisting system is written as
\begin{equation}\label{nlprsys}
\left\{
\begin{array}{ccl}
\dot{x}_{1}&\hspace{0pt}=&\hspace{0pt}\mbox{\boldmath $A$}_{1}x_{1}+
\mbox{\boldmath $L$}_{1}h_{2}(x_{2},x_{1})+\mbox{\boldmath $B$}_{1}u_{1}\vspace{-1mm}\\
\dot{x}_{2}&\hspace{0pt}=&\hspace{0pt}f_{2}(x_{2},x_{1})\vspace{-1.5mm}\\
y_{1}&\hspace{0pt}=&\hspace{0pt}\mbox{\boldmath $C$}_{1}x_{1}.
\end{array}
\right.
\end{equation}
Note that if (\ref{nlsys}) is a static nonlinear map, i.e., the dynamics of $x_{2}$ is empty and $\gamma_{2}=h_{2}(x_{1})$, the preexisting system (\ref{nlprsys}) can be regarded as a Lur'e system.
Assuming that (\ref{nlprsys}) is stable (i.e., globally input-to-state stable \citep{khalil1996nonlinear}), we can design a retrofit controller $\Pi_{1}$ in (\ref{retform}) such that the resultant closed-loop system is stable (i.e., globally asymptotically stable).
This is done by designing a local output feedback controller for the linear upstream dynamics (\ref{hrup}).


\section{Retrofit Control without Interconnection Signal Measurement}\label{secgrc}

\subsection{Problem Formulation}\label{secpf2}

Consider the preexisting system $\Sigma$ in (\ref{prsys}).
The objective of this section is to remove the assumption of the measurability of the interconnection signal $\gamma_{2}$ for the retrofit controller.
More specifically, the assumptions are listed as follows.

\begin{assumption}\label{assump2}\normalfont
For the preexisting system $\Sigma$ in (\ref{prsys}), the same assumptions \textbf{(i)} and \textbf{(ii)} as those in Assumption~\ref{assump} are made with 
\begin{description}
\item[(iii)] For the implementation of a retrofit controller, the measurement output signal $y_{1}$ is given by $y_{1}=x_{1}$, whereas the interconnection signal $\gamma_{2}$ is not measurable.
\end{description}
\end{assumption}

As compared with Assumption~\ref{assump}, the assumption on the measurability of $\gamma_{2}$ is removed while the availability of state feedback control is assumed for $\Sigma_{1}$.
We address the following retrofit controller design problem.

\begin{problem}\label{prob2}\normalfont
Consider the preexisting system $\Sigma$ in (\ref{prsys}) with the initial condition (\ref{stdef}).
Under Assumption~\ref{assump2}, find a retrofit controller of the form
\begin{equation}\label{retform2}
\Pi_{1}^{\prime}:u_{1}=\mathcal{K}_{1}^{\prime}(x_{1}),
\end{equation}
where $\mathcal{K}_{1}^{\prime}$ denotes a dynamical map, such that the same requirements \textbf{(A)} and \textbf{(B)} as those in Problem~\ref{prob1} are satisfied.
\end{problem}

\subsection{Solution}\label{secsol2}

To give a solution to Problem~\ref{prob2}, we introduce a parameterized version of hierarchical state-space expansion.
This parameterization plays an important role in the subsequent arguments.
As a generalization of Lemma~\ref{lemhr}, we state the following fact.

\begin{lemma}\label{lemhrp}\normalfont
Let $\mbox{\boldmath $P$}_{1}\in \mathbb{R}^{n_{1}\times\hat{n}_{1}}$ and $\mbox{\boldmath $P$}_{1}^{\dagger}\in \mathbb{R}^{\hat{n}_{1}\times n_{1}}$ denote a left invertible matrix and its left inverse, respectively.
For the preexisting system $\Sigma$ in (\ref{prsys}), consider the cascade interconnection system whose upstream subsystem is given by
\begin{subequations}\label{hrexp}
\begin{equation}\label{hrupp}
\dot{\hat{\xi}}_{1}=\mbox{\boldmath $P$}_{1}^{\dagger}\mbox{\boldmath $A$}_{1}\mbox{\boldmath $P$}_{1}\hat{\xi}_{1}+
\mbox{\boldmath $P$}_{1}^{\dagger}\mbox{\boldmath $B$}_{1}u_{1},
\end{equation}
which is $\hat{n}_{1}$-dimensional, and downstream subsystem is given by
\begin{equation}\label{hrdnp}
\left[\hspace{0pt}
\begin{array}{cc}
\dot{\xi}_{1}\\
\dot{\xi}_{2}
\end{array}
\hspace{0pt}
\right]
=
\left[\hspace{0pt}
\begin{array}{cc}
\mbox{\boldmath $A$}_{1}&\hspace{0pt}\mbox{\boldmath $L$}_{1}{\mathit \Gamma}_{2}\\
 L_{2}{\mathit \Gamma}_{1}&\hspace{0pt}A_{2}
\end{array}
\hspace{0pt}
\right]
\left[\hspace{0pt}
\begin{array}{cc}
\xi_{1}\\
\xi_{2}
\end{array}
\hspace{0pt}
\right]
+
\left[\hspace{0pt}
\begin{array}{cc}
\overline{\mbox{\boldmath $P$}}_{1}\overline{\mbox{\boldmath $P$}}_{1}^{\dagger}\mbox{\boldmath $A$}_{1}\\
 L_{2}{\mathit \Gamma}_{1}
\end{array}
\hspace{0pt}
\right]\mbox{\boldmath $P$}_{1}\hat{\xi}_{1},
\end{equation}
\end{subequations}
which is $(n_{1}+n_{2})$-dimensional, 
where a left invertible matrix $\overline{\mbox{\boldmath $P$}}_{1}\in \mathbb{R}^{n_{1}\times(n_{1}-\hat{n}_{1})}$ and its left inverse 
$\overline{\mbox{\boldmath $P$}}_{1}^{\dagger}\in \mathbb{R}^{(n_{1}-\hat{n}_{1})\times n_{1}}$ are given such that
\begin{equation}\label{projs}
\mbox{\boldmath $P$}_{1}\mbox{\boldmath $P$}_{1}^{\dagger}
+\overline{\mbox{\boldmath $P$}}_{1}\overline{\mbox{\boldmath $P$}}_{1}^{\dagger}=I.
\end{equation}
If $\mbox{\boldmath $P$}_{1}$ satisfies
\begin{equation}\label{imBcon}
{\rm im}\ \!\mbox{\boldmath $B$}_{1}\subseteq{\rm im}\ \!\mbox{\boldmath $P$}_{1},
\end{equation}
then it follows that
\begin{equation}\label{stateeqp}
x_{1}(t)=\xi_{1}(t)+\mbox{\boldmath $P$}_{1}\hat{\xi}_{1}(t),\quad x_{2}(t)=\xi_{2}(t),\quad\forall t\geq 0
\end{equation}
for any external input signal $u_{1}$, provided that (\ref{stateeqp}) is satisfied at the initial time $t=0$.
\end{lemma}

Note that $\mbox{\boldmath $P$}_{1}\mbox{\boldmath $P$}_{1}^{\dagger}\mbox{\boldmath $B$}_{1}=\mbox{\boldmath $B$}_{1}$ if (\ref{imBcon}) holds.
Thus, the claim can be proved by summing (\ref{hrupp}) multiplied by $\mbox{\boldmath $P$}_{1}$ and (\ref{hrdnp}).
The hierarchical realization (\ref{hrexp}) involves $\mbox{\boldmath $P$}_{1}$ and $\mbox{\boldmath $P$}_{1}^{\dagger}$ as free parameters.
The product $\overline{\mbox{\boldmath $P$}}_{1}\overline{\mbox{\boldmath $P$}}_{1}^{\dagger}$ is determined by these parameters according to (\ref{projs}).
Clearly, if we take both $\mbox{\boldmath $P$}_{1}$ and $\mbox{\boldmath $P$}_{1}^{\dagger}$ as the identity, then (\ref{hrexp}) coincides with (\ref{hrex}).
Note that the upstream dynamics (\ref{hrupp}) is a low-dimensional approximate model of (\ref{hrup}) obtained by an oblique projection \citep{antoulas2005approximation}.

For consistency with (\ref{stdef}) and (\ref{stateeqp}), we describe the initial condition of the hierarchical realization (\ref{hrexp}) as
\begin{equation}\label{ginicon}
\hspace{2pt}
\hat{\xi}_{1}(0)=\mbox{\boldmath $P$}_{1}^{\dagger}(\delta_{0}-\zeta_{0}),\quad
\left[\hspace{0pt}
\begin{array}{cc}
\xi_{1}(0)\\
\xi_{2}(0)
\end{array}
\hspace{0pt}
\right]\!=\!
\left[\hspace{0pt}
\begin{array}{cc}
\scriptsize{
\overline{\mbox{\boldmath $P$}}_{1}\overline{\mbox{\boldmath $P$}}_{1}^{\dagger}\delta_{0}+
\mbox{\boldmath $P$}_{1}\mbox{\boldmath $P$}_{1}^{\dagger}\zeta_{0}
}\\
0
\end{array}
\right]
\hspace{-12pt}
\end{equation}
where $\zeta_{0}\in \mathbb{R}^{n_{1}}$ is an arbitrary parameter.
Based on this parameterized hierarchical realization, let us consider the design of a local state feedback controller.
For the upstream dynamics (\ref{hrupp}), a local state feedback controller
\begin{equation}\label{stfbcon}
u_{1}=\mbox{\boldmath $\hat{K}$}_{1}\hat{\xi}_{1}
\end{equation}
is designed such that the closed-loop dynamics
\begin{equation}\label{stapsys}
\dot{\hat{\xi}}_{1}=
(\mbox{\boldmath $P$}_{1}^{\dagger}\mbox{\boldmath $A$}_{1}\mbox{\boldmath $P$}_{1}
+\mbox{\boldmath $P$}_{1}^{\dagger}\mbox{\boldmath $B$}_{1}\mbox{\boldmath $\hat{K}$}_{1})\hat{\xi}_{1}
\end{equation}
is internally stable, and the control performance specification
\begin{equation}\label{gxihprf}
\|\hat{\xi}_{1}\|_{\mathcal{L}_{2}}\leq\epsilon_{1},\quad\forall\hat{\xi}_{1}(0)\in\hat{\mathcal{B}}
\end{equation}
is satisfied for a given tolerance $\epsilon_{1}>0$, where
\[
\hat{\mathcal{B}}:=\{\mbox{\boldmath $P$}_{1}^{\dagger}\delta_{0}\in \mathbb{R}^{\hat{n}_{1}}:\delta_{0}\in \mathcal{B}\}.
\]
Then, Lemma~\ref{lemspec} can be generalized as follows.

\begin{lemma}\label{lemspec2}\normalfont
For the hierarchical realization (\ref{hrexp}), consider the local state feedback controller (\ref{stfbcon}).
Under Assumption~\ref{assump2}~\textbf{(i)}, the closed-loop system composed of (\ref{hrexp}) and (\ref{stfbcon}) is internally stable if and only if the closed-loop dynamics (\ref{stapsys}) is internally stable.
Furthermore, let
\begin{equation}\label{Gis2}
G_{i}^{\prime}(s):=E_{i}^{{\sf T}}(sI-A)^{-1}
\bigl\{E_{1}\overline{\mbox{\boldmath $P$}}_{1}\overline{\mbox{\boldmath $P$}}_{1}^{\dagger}\mbox{\boldmath $A$}_{1}
+E_{2}L_{2}{\mathit \Gamma}_{1}\bigr\}
\end{equation}
denote the transfer matrix from $\mbox{\boldmath $P$}_{1}\hat{\xi}_{1}$ to $\xi_{i}$ of the downstream dynamics (\ref{hrdnp}), where $A$ is defined as in (\ref{asclsta}) and $E_{1}$,  $E_{2}$ are defined as in (\ref{defE}).
If (\ref{gxihprf}) holds for the closed-loop dynamics (\ref{stapsys}), then
\begin{equation}\label{prfbnd2}
\hspace{0pt}
\begin{array}{rcl}
\|\xi_{1}\!+\!\mbox{\boldmath $P$}_{1}\hat{\xi}_{1}\|_{\mathcal{L}_{2}}&\hspace{-2pt}\leq&\hspace{-2pt}\alpha_{1}^{\prime}
(1\!+\!\|\zeta_{0}\|)\epsilon_{1}\!+\!\beta_{1}^{\prime}(\delta_{0},\zeta_{0}),\vspace{0pt}\\
\|\xi_{2}\|_{\mathcal{L}_{2}}&\hspace{-2pt}\leq&\hspace{-2pt}\alpha_{2}^{\prime}(1\!+\!\|\zeta_{0}\|)\epsilon_{1}\!+\!\beta_{2}^{\prime}(\delta_{0},\zeta_{0}),
\end{array}
\quad\forall\delta_{0}\in \mathcal{B}
\end{equation}
with the initial condition (\ref{ginicon}), where the nonnegative constants
\begin{equation}\label{consts}
\alpha_{1}^{\prime}:=\|(G_{1}^{\prime}+I)\mbox{\boldmath $P$}_{1}\|_{\mathcal{H}_{\infty}},\quad
\alpha_{2}^{\prime}:=\|G_{2}^{\prime}\mbox{\boldmath $P$}_{1}\|_{\mathcal{H}_{\infty}},
\end{equation}
and the nonnegative functions
\begin{equation}\label{betas}
\beta_{i}^{\prime}(\delta_{0},\zeta_{0}):=\bigl\|E_{i}^{{\sf T}}e^{At}E_{1}
(\overline{\mbox{\boldmath $P$}}_{1}\overline{\mbox{\boldmath $P$}}_{1}^{\dagger}\delta_{0}+
\mbox{\boldmath $P$}_{1}\mbox{\boldmath $P$}_{1}^{\dagger}\zeta_{0})
\bigr\|_{\mathcal{L}_{2}}
\end{equation}
are independent of the selection of the feedback gain $\mbox{\boldmath $\hat{K}$}_{1}$ in (\ref{stfbcon}).
\end{lemma}

Owing to the cascade structure of the hierarchical realization, this claim can be proved in a similar manner to the proof of Lemma~\ref{lemspec}.
Let us consider selecting $\zeta_{0}$ as in (\ref{zeta0}).
Then, we discuss how to implement the local state feedback controller (\ref{stfbcon}) for the original realization $\Sigma$ in (\ref{prsys}).
Note that $\hat{\xi}_{1}$ is equal to $\mbox{\boldmath $P$}_{1}^{\dagger}x_{1}-\mbox{\boldmath $P$}_{1}^{\dagger}\xi_{1}$ owing to (\ref{stateeqp}).
To generate $\mbox{\boldmath $P$}_{1}^{\dagger}\xi_{1}$, we implement an output rectifier that achieves
\begin{equation}\label{gxhat1}
\hat{x}_{1}(t)=\mbox{\boldmath $P$}_{1}^{\dagger}\xi_{1}(t),\quad\forall t\geq 0,
\end{equation}
where $\hat{x}_{1}$ denotes the state of the output rectifier.
Considering (\ref{stateeqp}) and (\ref{gxhat1}) as the coordinate transformation from the hierarchical realization to the original realization, whose inverse is given by
\begin{equation}\label{ctrans2}
\left[\hspace{0pt}
\begin{array}{cc}
\xi_{1}\\
\xi_{2}\\
\hat{\xi}_{1}
\end{array}
\hspace{0pt}
\right]
=
\left[\hspace{0pt}
\begin{array}{ccc}
\overline{\mbox{\boldmath $P$}}_{1}\overline{\mbox{\boldmath $P$}}_{1}^{\dagger}&0&\mbox{\boldmath $P$}_{1}\\
0&I&0\\
\mbox{\boldmath $P$}_{1}^{\dagger}&0&-I
\end{array}
\hspace{0pt}
\right]
\left[\hspace{0pt}
\begin{array}{cc}
x_{1}\\
x_{2}\\
\hat{x}_{1}
\end{array}
\hspace{0pt}
\right],
\end{equation}
we verify that a realization of the output rectifier is given by
\begin{equation}\label{dyncomp2}
\left\{
\begin{array}{ccl}
\dot{\hat{x}}_{1}&\hspace{0pt}=&\hspace{0pt}\mbox{\boldmath $P$}_{1}^{\dagger}\mbox{\boldmath $A$}_{1}\mbox{\boldmath $P$}_{1}\hat{x}_{1}+\mbox{\boldmath $P$}_{1}^{\dagger}\mbox{\boldmath $A$}_{1}
\overline{\mbox{\boldmath $P$}}_{1}\overline{\mbox{\boldmath $P$}}_{1}^{\dagger}x_{1}+
\mbox{\boldmath $P$}_{1}^{\dagger}\mbox{\boldmath $L$}_{1}\gamma_{2}\vspace{-1mm}\\
\hat{y}_{1}&\hspace{0pt}=&\hspace{0pt}\mbox{\boldmath $P$}_{1}^{\dagger}x_{1}-\hat{x}_{1},
\end{array}
\right.
\end{equation}
where the initial condition is determined to be (\ref{inixh}) because of (\ref{zeta0}).
This initial condition is actually consistent with (\ref{ginicon}) and (\ref{gxhat1}) because of $\mbox{\boldmath $P$}_{1}^{\dagger}\overline{\mbox{\boldmath $P$}}_{1}=0$, which comes from the fact that (\ref{projs}) implies 
\[
\left[\hspace{0pt}
\begin{array}{cc}
\mbox{\boldmath $P$}_{1}&
\overline{\mbox{\boldmath $P$}}_{1}
\end{array}
\hspace{0pt}
\right]
\left[\hspace{0pt}
\begin{array}{cc}
\mbox{\boldmath $P$}_{1}^{\dagger}\\
\overline{\mbox{\boldmath $P$}}_{1}^{\dagger}
\end{array}
\hspace{0pt}
\right]=I\quad \Longleftrightarrow \quad
\left[\hspace{0pt}
\begin{array}{cc}
\mbox{\boldmath $P$}_{1}^{\dagger}\\
\overline{\mbox{\boldmath $P$}}_{1}^{\dagger}
\end{array}
\hspace{0pt}
\right]\left[\hspace{0pt}
\begin{array}{cc}
\mbox{\boldmath $P$}_{1}&
\overline{\mbox{\boldmath $P$}}_{1}
\end{array}
\hspace{0pt}
\right]=I.
\]
Note that (\ref{ctrans2}) and (\ref{dyncomp2}) correspond to the generalization of (\ref{ctrans}) and (\ref{dyncomp}), respectively.
However, in the output rectifier (\ref{dyncomp2}), note the appearance of a term containing the interconnection signal $\gamma_{2}$.
To remove this term, we use the remaining degree of freedom to assign the kernel of $\mbox{\boldmath $P$}_{1}^{\dagger}$.
To this end, we state the following fact.

\begin{lemma}\label{lemPiff}\normalfont
Consider the subsystem $\Sigma_{1}$ in (\ref{sys1}).
There exist a left invertible matrix $\mbox{\boldmath $P$}_{1}$ and its left inverse $\mbox{\boldmath $P$}_{1}^{\dagger}$ such that
\begin{equation}\label{spcon}
{\rm im}\ \!\mbox{\boldmath $B$}_{1}\subseteq{\rm im}\ \!\mbox{\boldmath $P$}_{1},\quad
{\rm im}\ \!\mbox{\boldmath $L$}_{1}\subseteq{\rm ker}\ \!\mbox{\boldmath $P$}_{1}^{\dagger}
\end{equation}
if and only if
\begin{equation}\label{BLcon}
{\rm im}\ \!\mbox{\boldmath $B$}_{1}\cap{\rm im}\ \!\mbox{\boldmath $L$}_{1}=\emptyset.
\end{equation}
\end{lemma}\vspace{-2mm}

\begin{pf}
We first prove the sufficiency, i.e., if (\ref{BLcon}) holds, then there exist $\mbox{\boldmath $P$}_{1}$ and $\mbox{\boldmath $P$}_{1}^{\dagger}$ such that (\ref{spcon}) holds.
As shown in Proposition~3.5.3 of \citep{bernstein2009matrix}, for any complementary subspaces $\mathcal{V}_{1}$ and $\mathcal{V}_{2}$, there exists the unique projection matrix $H_{1}$ onto $\mathcal{V}_{1}$ along $\mathcal{V}_{2}$.
A realization of this matrix is
\begin{equation}\label{projH}
H_{1}=V_{1}(V_{2}^{\sf T}V_{1})^{-1}V_{2}^{\sf T},\quad
\left\{
\begin{array}{ccl}
\mathcal{V}_{1}&\hspace{-0pt}=&\hspace{-0pt}{\rm im}\ \!V_{1}\vspace{-1mm}\\
\mathcal{V}_{2}&\hspace{-0pt}=&{\rm ker}\ \!V_{2}^{\sf T}.
\end{array}
\right.
\end{equation}
Because (\ref{BLcon}) implies that the column vectors of $\mbox{\boldmath $B$}_{1}$ and $\mbox{\boldmath $L$}_{1}$ are linearly independent, the complementary subspaces such that 
${\rm im}\ \!\mbox{\boldmath $B$}_{1}\subseteq \mathcal{V}_{1}$ and 
${\rm im}\ \!\mbox{\boldmath $L$}_{1}\subseteq \mathcal{V}_{2}$ can be selected.
Thus, the selection of
\[
\mbox{\boldmath $P$}_{1}=V_{1}
(V_{2}^{\sf T}V_{1})^{-1},
\quad
\mbox{\boldmath $P$}_{1}^{\dagger}=V_{2}^{\sf T}
\]
satisfies (\ref{BLcon}).
This proves the sufficiency.

Next, to prove the necessity, we consider the contraposition.
Namely, if (\ref{BLcon}) does not hold, i.e., if there exists some vector $v$ such that
\[
v\in{\rm im}\ \!\mbox{\boldmath $B$}_{1},\quad v\in{\rm im}\ \!\mbox{\boldmath $L$}_{1},
\]
then there exist no $\mbox{\boldmath $P$}_{1}$ and $\mbox{\boldmath $P$}_{1}^{\dagger}$ such that (\ref{spcon}) holds.
Equivalently, there is no projection matrix $H_{1}$ in (\ref{projH}) onto the image of $\mbox{\boldmath $P$}_{1}$ along the kernel of $\mbox{\boldmath $P$}_{1}^{\dagger}$, whose realization is $\mbox{\boldmath $P$}_{1}\mbox{\boldmath $P$}_{1}^{\dagger}$, such that (\ref{spcon}) holds.
Note that $H_{1}v=v$ for $v\in{\rm im}\ \!\mbox{\boldmath $B$}_{1}$, while $H_{1}v=0$ for $v\in{\rm im}\ \!\mbox{\boldmath $L$}_{1}$.
They are contradictory.
This proves the necessity.\hfill$\square $
\end{pf}\vspace{-2mm}

Lemma~\ref{lemPiff} implies that we can always find a pair of $\mbox{\boldmath $P$}_{1}$ and $\mbox{\boldmath $P$}_{1}^{\dagger}$ such that (\ref{spcon}) holds, provided that the column vectors of $\mbox{\boldmath $B$}_{1}$ and $\mbox{\boldmath $L$}_{1}$ are linearly independent as described in (\ref{BLcon}).
The image condition for $\mbox{\boldmath $P$}_{1}$ in (\ref{spcon}) is necessary to make the hierarchical state-space expansion valid as shown in Lemma~\ref{lemhrp}.
The kernel condition for $\mbox{\boldmath $P$}_{1}^{\dagger}$ is used to remove the term containing $\mbox{\boldmath $P$}_{1}^{\dagger}\mbox{\boldmath $L$}_{1}$ in (\ref{dyncomp2}).
Note that (\ref{BLcon}) is generally a mild condition that simply implies the control input port and  interconnection input port are not exactly equal.
In conclusion, a solution to Problem~\ref{prob2} can be formally stated as follows.

\begin{theorem}\label{thmcocon}\normalfont
Under Assumption~\ref{assump2}~\textbf{(i)} with the condition (\ref{BLcon}), consider the preexisting system $\Sigma$ in (\ref{prsys}) with the initial condition (\ref{stdef}).
Let $\mbox{\boldmath $P$}_{1}$ and $\mbox{\boldmath $P$}_{1}^{\dagger}$ be a left invertible matrix and its left inverse such that (\ref{spcon}) holds.
Then, for any local state feedback controller (\ref{stfbcon}) such that the closed-loop dynamics (\ref{stapsys}) is internally stable and (\ref{gxihprf}) holds, the entire closed-loop system composed of (\ref{prsys}) and
\begin{equation}\label{rcstar}
\Pi_{1}^{\prime}:\left\{
\begin{array}{ccl}
\dot{\hat{x}}_{1}&\hspace{0pt}=&\hspace{0pt}\mbox{\boldmath $P$}_{1}^{\dagger}\mbox{\boldmath $A$}_{1}\mbox{\boldmath $P$}_{1}\hat{x}_{1}+\mbox{\boldmath $P$}_{1}^{\dagger}\mbox{\boldmath $A$}_{1}
\overline{\mbox{\boldmath $P$}}_{1}\overline{\mbox{\boldmath $P$}}_{1}^{\dagger}x_{1}\vspace{-1mm}\\
u_{1}&\hspace{0pt}=&\hspace{0pt}\mbox{\boldmath $\hat{K}$}_{1}(\mbox{\boldmath $P$}_{1}^{\dagger}x_{1}-\hat{x}_{1})
\end{array}
\right.
\end{equation}
with the initial condition (\ref{inixh}) is internally stable and
\begin{equation}\label{eprfbnd2}
\|x_{i}\|_{\mathcal{L}_{2}}\leq\alpha_{i}^{\prime}\epsilon_{1}+\beta_{i}(\overline{\mbox{\boldmath $P$}}_{1}\overline{\mbox{\boldmath $P$}}_{1}^{\dagger}\delta_{0}),\quad
\forall\delta_{0}\in \mathcal{B}
\end{equation}
for each $i\in\{1,2\}$, where $\alpha_{i}^{\prime}$ in (\ref{consts}) and $\beta_{i}$ in (\ref{betai}) are independent of the local controller design of (\ref{stfbcon}) provided that $\mbox{\boldmath $P$}_{1}$ and $\mbox{\boldmath $P$}_{1}^{\dagger}$ are determined before the local controller is designed.
\end{theorem}

As shown in Theorem~\ref{thmcocon}, the resultant retrofit controller $\Pi_{1}^{\prime}$ in (\ref{rcstar}) is formed as the cascade interconnection of the local state feedback controller (\ref{stfbcon}) and the $\hat{n}_{1}$-dimensional output rectifier (\ref{dyncomp2}) from which the term containing the interconnection signal $\gamma_{2}$ has been removed.
Note that the remarks in Sections~\ref{secsub1}, \ref{secsmrc}, and \ref{secgnc} also apply.
The retrofit controller $\Pi_{1}^{\prime}$ can be regarded as a dynamical controller with full state information of $\Sigma_{1}$.
This can be seen from the fact that $\overline{\mbox{\boldmath $P$}}_{1}\overline{\mbox{\boldmath $P$}}_{1}^{\dagger}x_{1}$ in the output rectifier corresponds to the projection of $x_{1}$ onto the kernel of $\mbox{\boldmath $P$}_{1}^{\dagger}$ along the kernel of $\overline{\mbox{\boldmath $P$}}_{1}^{\dagger}$.
In contrast, $\mbox{\boldmath $P$}_{1}^{\dagger}x_{1}$ in the local state feedback controller eliminates the component of $x_{1}$ in the kernel of $\mbox{\boldmath $P$}_{1}^{\dagger}$, which is neglected in the local controller design with the projected model (\ref{hrupp}).
They are actually complementary.

\section{Numerical Examples}\label{secnumex}

\subsection{Frequency Control for Power Systems}\label{secnps}

\begin{figure}[t]
\begin{center}
\includegraphics[width=60mm]{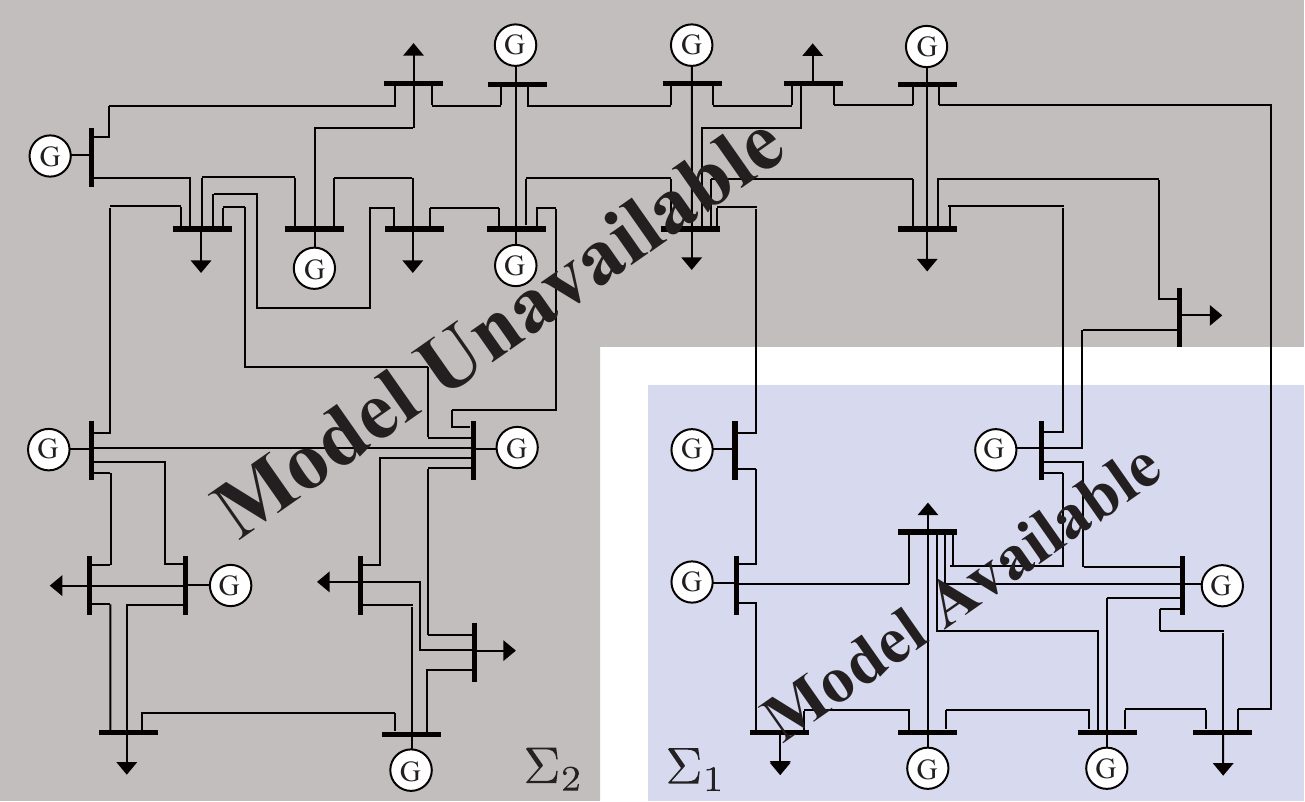}
\end{center}
\vspace{0pt}
\caption{\scriptsize
Power network model composed of generators and loads. Generators are denoted by ``G'' and loads are denoted by ``$\downarrow$.'' }
\label{figpowernet}
\end{figure}

In this subsection, we demonstrate the significance of the theory in Section~\ref{secfrc}.
The theory in Section~\ref{secgrc} will be used in Section~\ref{seccac}.
We consider a power network model composed of 16 generators and 14 loads, where the network structure is as depicted in Fig.~\ref{figpowernet}.
According to \citep{ilic1996hierarchical,chakrabortty2011wide}, the dynamics of each generator is described as a rotary appliance
\begin{equation}\label{rotar}
\dot{\theta}_{i}=\omega_{i},\quad m_{i}\dot{\omega}_{i}+d_{i}\omega_{i}+f_{i}+e_{i}=0
\end{equation}
with a second order governor
\begin{equation}\label{goven}
\tau_{i}\dot{f}_{i}=-f_{i}+p_{i},\quad\tau^{\prime}_{i}\dot{p}_{i}=-\kappa_{i}p_{i}+\omega_{i}+v_{i},
\end{equation}
where $\theta_{i}$ and $\omega_{i}$ denote the phase angle and frequency, $f_{i}$ and $e_{i}$ denote the mechanical torque from the governor and the electric torque from other appliances, $p_{i}$ denotes the valve position, and $v_{i}$ denotes the control input signal to the governor.
In a similar way, we describe the load dynamics as the rotary appliance (\ref{rotar}) without the mechanical torque term $f_{i}$.
Each inertia constant $m_{i}\in[2,10]$ and damping constant $d_{i}\in[0.001,0.1]$ for the generators and loads is randomly selected.
We set the turbine constant $\tau_{i}=0.002$, the governor time constant $\tau_{i}^{\prime}=1$, and the droop constant $\kappa_{i}=0.1$ for all generators.
The interconnection between the generators and loads can be represented as
\begin{equation}
\textstyle
e_{i}=\sum_{j\in \mathcal{N}_{i}}Y_{i,j}(\theta_{j}-\theta_{i})
\end{equation}
where $\mathcal{N}_{i}$ denotes the index set associated with the neighborhood of the $i$th appliance and $Y_{i,j}$ denotes the admittance between the $i$th and $j$th appliances.
Each admittance value is selected from $[1,40]$.
In the following, we assume that all generator and load variables are defined in terms of their deviation from desirable equilibria.

\begin{figure*}[t]
\begin{center}
\includegraphics[width=175mm]{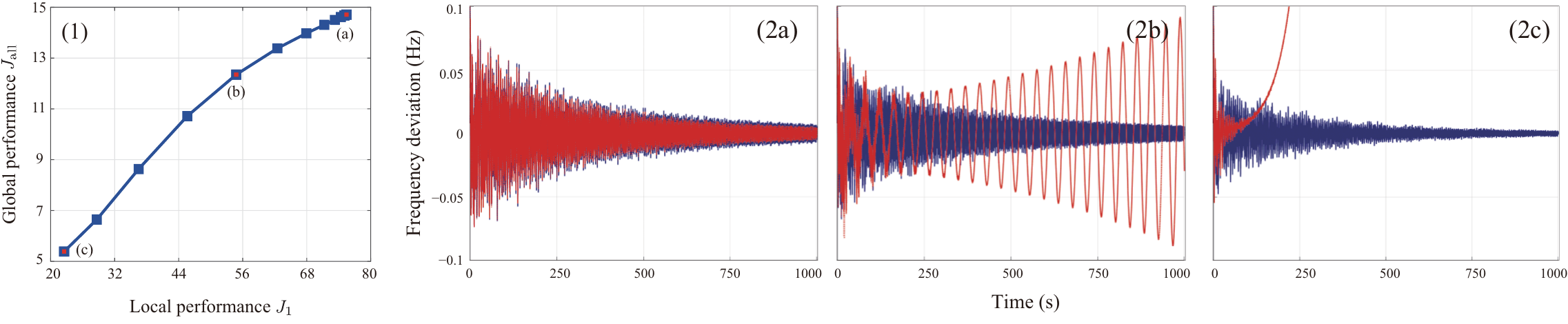}
\end{center}
\vspace{-8pt}
\caption{\scriptsize(1) Global performance versus local performance. (2a)-(2c) Frequency deviation trajectories of the appliances in the first subsystem.}
\label{figlmhgain}
\end{figure*}

We consider implementing a retrofit controller for the subsystem $\Sigma_{1}$ in Fig.~\ref{figpowernet}, whose  system model is assumed to be available.
For the output signals, we assume that the frequencies and phase angles of all generators in $\Sigma_{1}$ are measurable.
In addition, the interconnection signal from $\Sigma_{2}$ is assumed to be measurable.
The retrofit controller is designed for $\Sigma_{1}$ as an observer-based state feedback controller in the form of (\ref{obsretro}), whose feedback gains $\mbox{\boldmath $F$}_{1}$ and $\mbox{\boldmath $H$}_{1}$ are determined for the isolated model of $\Sigma_{1}$ based on the linear quadratic regulator design technique.

For the subsequent discussion, let us define the global and local control performance measures as
\[
J_{{\rm all}}=\sup_{\omega(0)\in \mathcal{U}}\|\omega\|_{\mathcal{L}_{2}},\quad
J_{1}=\sup_{\hat{\omega}_{1}(0)\in \mathcal{U}}\|\hat{\omega}_{1}\|_{\mathcal{L}_{2}},
\]
where $\mathcal{U}$ denotes the set of vectors having the unit norm, $\omega$ denotes the frequency deviation vector for all appliances, and $\hat{\omega}_{1}$ denotes the frequency deviation vector of the appliances in $\Sigma_{1}$ when the interconnection with $\Sigma_{2}$ is neglected.
Note that the value of $J_{1}$ corresponds to that of $\epsilon_{1}$ in (\ref{xihprf}).
By varying the quadratic weights for the controller design, we plot the resultant values of $J_{{\rm all}}$ versus the values of $J_{1}$ in Fig.~\ref{figlmhgain}(1), where (a), (b), and (c) correspond to low-gain, medium-gain, and high-gain retrofit controllers, respectively.
From this figure, we see that the global control performance improves as the local performance improves.

The resultant frequency deviation trajectories of the appliances in $\Sigma_{1}$ are plotted in the right of Figs.~\ref{figlmhgain} (2a)-(2c), where the initial frequency deviation of each appliance 
in $\Sigma_{1}$, corresponding to $\delta_{0}$ in (\ref{stdef}), is randomly selected from $[0,0.2]$.
Each subfigure corresponds to the indication of (a)-(c) in Fig.~\ref{figlmhgain}(1).
The blue solid lines correspond to the case of a retrofit controller with the output rectifier, whereas the red dotted lines correspond to the case with no output rectifier.
This result shows that the output rectifier involved in the retrofit controller plays a significant role in ensuring whole-system stability, even when the simple implementation of medium-gain, and high-gain local controllers without the output rectifier induces system instability.

\subsection{Vehicle Platoon Control for Collision Avoidance}\label{seccac}

\begin{figure}[t]
\begin{center}
\includegraphics[width=80mm]{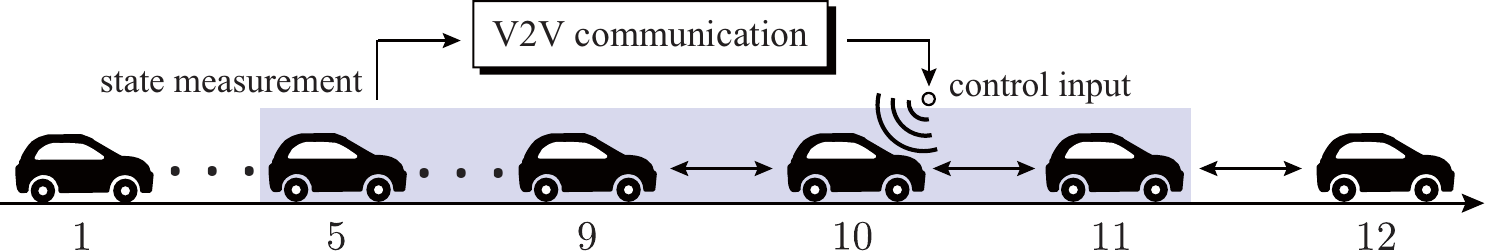}
\end{center}
\vspace{-4pt}
\caption{\scriptsize{
Vehicle platoon control.
The controller measures the states of neighboring vehicles though a vehicle-to-vehicle (V2V) communication.}}
\label{figplatoon}
\end{figure}

We demonstrate the significance of the theory in Section~\ref{secgrc} with the nonlinear generalization in Section~\ref{secgnc}.
Let us consider the platoon of 12 vehicles depicted as in Fig.~\ref{figplatoon}, where the labels are assigned from the headmost vehicle in descending order.
Supposing that the velocity of each vehicle is operated by a driver, for $i\in\{1,\ldots,12\}$, we model the $i$th vehicle dynamics \citep{hayakawa1998theory} as
\begin{equation}\label{dynpltn}
\left\{
\begin{array}{ccl}
\dot{p}_{i}&\hspace{0pt}=&\hspace{0pt}v_{i}\vspace{-1.5mm}\\
\dot{v}_{i}&\hspace{0pt}=&\hspace{0pt}\kappa\left\{f(p_{i+1}-p_{i})g(p_{i}-p_{i-1})-v_{i}\right\}+w_{i},
\end{array}
\right.
\end{equation}
where $p_{i}$ and $v_{i}$ denote the position and velocity,  $\kappa$ denotes a positive constant representing sensitivity to the forward and backward vehicles, and $w_{i}$ denotes the external input signal.
We set the sensitivity constant as $\kappa=0.06$ and the nonlinear functions as
\[
\begin{array}{ccl}
f(x)&\hspace{0pt}=&\hspace{0pt}\tanh(x-2)+\tanh(2),\vspace{-1mm}\\
g(x)&\hspace{0pt}=&\hspace{0pt}1+5\left\{1-\tanh(3x-6.3)\right\},
\end{array}
\]
where $f$ is monotone increasing and bounded, and $g$ is monotone decreasing, bounded, and $g(\infty)=1$.
These functions represent a driver operating property so as to avoid a collision with the forward and backward vehicles.
In particular, $g$ can be regarded as a scaling factor for the acceleration because its range of values is greater than or equal to $1$.

Assuming that the desired inter-vehicle distance, denoted by $\Delta p^{*}$, is $2.7$, we regard $p_{13}$ as $p_{12}+\Delta p^{*}$ and $p_{0}$ as $p_{1}-\Delta p^{*}$.
As shown in \citep{hayakawa1998theory}, the equilibrium trajectory of (\ref{dynpltn}) without $w_{i}$ is given by
\begin{equation}\label{eqtra}
p_{i}(t)=i\Delta p^{*}+\overline{v}t,\quad v_{i}(t)=\overline{v},\quad i\in\{1,\ldots,12\},
\end{equation}
where $\overline{v}:=f(\Delta p^{*})g(\Delta p^{*})$, and this is stable as long as $\kappa$ is above than a certain threshold.
Hereafter, we assume the stability of this equilibrium trajectory.

For the vehicle platoon model (\ref{dynpltn}), we consider the design of a retrofit controller that works inside a vehicle to prevent collisions caused by sudden braking.
In particular, we suppose that the retrofit controller is implemented in Vehicle~10, i.e., all control inputs $w_{i}$ other than $w_{10}$ are zero, and it can measure the positions and velocities of Vehicles~5--11 through V2V communication, as depicted in Fig.~\ref{figplatoon}.
This means that the retrofit controller is designed as a state feedback controller that injects the control input to Vehicle~10 while measuring the states of Vehicles~5--11.

Because the vehicle platoon model (\ref{dynpltn}) is a nonlinear system, we consider the subsystem $\Sigma_{1}$ in (\ref{sys1}) as a linear approximation of the dynamics corresponding to Vehicles~5--11, which is a 14-dimensional system.
The linear dynamics is obtained by the linearization around the stable equilibrium trajectory, and can be represented as
\[
\dot{x}_{1}=\mbox{\boldmath $A$}_{1}x_{1}+
\mbox{\boldmath $L$}_{1}\gamma_{2}+\gamma_{1}+\mbox{\boldmath $B$}_{1}u_{1},
\]
where $\gamma_{2}$ corresponds to the interconnection signal from Vehicles~4 and 12, $\gamma_{1}$ corresponds to the nonlinear term neglected though the linearization, and $u_{1}$ corresponds to $w_{10}$.
On the other hand, the subsystem $\Sigma_{2}$, given as a nonlinear system in (\ref{nlsys}),  is composed of the static nonlinear term of Vehicles 5--11, and the nonlinear dynamics of the remaining vehicles, which is a 10-dimensional system.
This can be represented as
\[
\dot{x}_{2}=f_{2}(x_{2},x_{1}),\quad
\gamma_{1}=\mbox{\boldmath $f$}_{1}(x_{1}),\quad
\gamma_{2}=h_{2}(x_{2})
\]
where $\gamma_{1}$ is measurable owing to the measurability of $x_{1}$ but $\gamma_{2}$ is not.
Note that the control input port is located at Vehicle~10, whereas the interconnection input ports are located at Vehicles~5 and 11.
This means that the condition (\ref{BLcon}) is satisfied, i.e., there exist $\mbox{\boldmath $P$}_{1}$ and $\mbox{\boldmath $P$}_{1}^{\dagger}$ such that (\ref{spcon}) holds.
In this case, the retrofit controller has the form
\begin{equation}\label{retex}
\left\{
\begin{array}{ccl}
\dot{\hat{x}}_{1}&\hspace{0pt}=&\hspace{0pt}\mbox{\boldmath $P$}_{1}^{\dagger}\mbox{\boldmath $A$}_{1}\mbox{\boldmath $P$}_{1}\hat{x}_{1}+\mbox{\boldmath $P$}_{1}^{\dagger}\mbox{\boldmath $f$}_{1}(x_{1})+\mbox{\boldmath $P$}_{1}^{\dagger}\mbox{\boldmath $A$}_{1}
\overline{\mbox{\boldmath $P$}}_{1}\overline{\mbox{\boldmath $P$}}_{1}^{\dagger}x_{1}\vspace{-1.5mm}\\
u_{1}&\hspace{0pt}=&\hspace{0pt}\mbox{\boldmath $\hat{K}$}_{1}(\mbox{\boldmath $P$}_{1}^{\dagger}x_{1}-\hat{x}_{1}).
\end{array}
\right.\hspace{-12pt}
\end{equation}
We first compare the controller design given by the linearization.
The dimension of the retrofit controller is taken as $\hat{n}_{1}=12$.
This is the maximal number such that (\ref{spcon}) holds because $\hat{n}_{1}$ must satisfy
\begin{equation}\label{dcbd}
{\rm rank}\ \!\mbox{\boldmath $B$}_{1}\leq\hat{n}_{1}\leq n_{1}-{\rm rank}\ \!\mbox{\boldmath $L$}_{1},
\end{equation}
where $n_{1}=14$ and ${\rm rank}\ \!\mbox{\boldmath $L$}_{1}=2$.
Based on the linear quadratic regulator design technique, we calculate the optimal feedback gain $\mbox{\boldmath $\hat{K}$}_{1}$ with respect to a quadratic cost function such that (\ref{stapsys}) exhibits desirable behavior.

\begin{figure*}[t]
\begin{center}
\includegraphics[width=170mm]{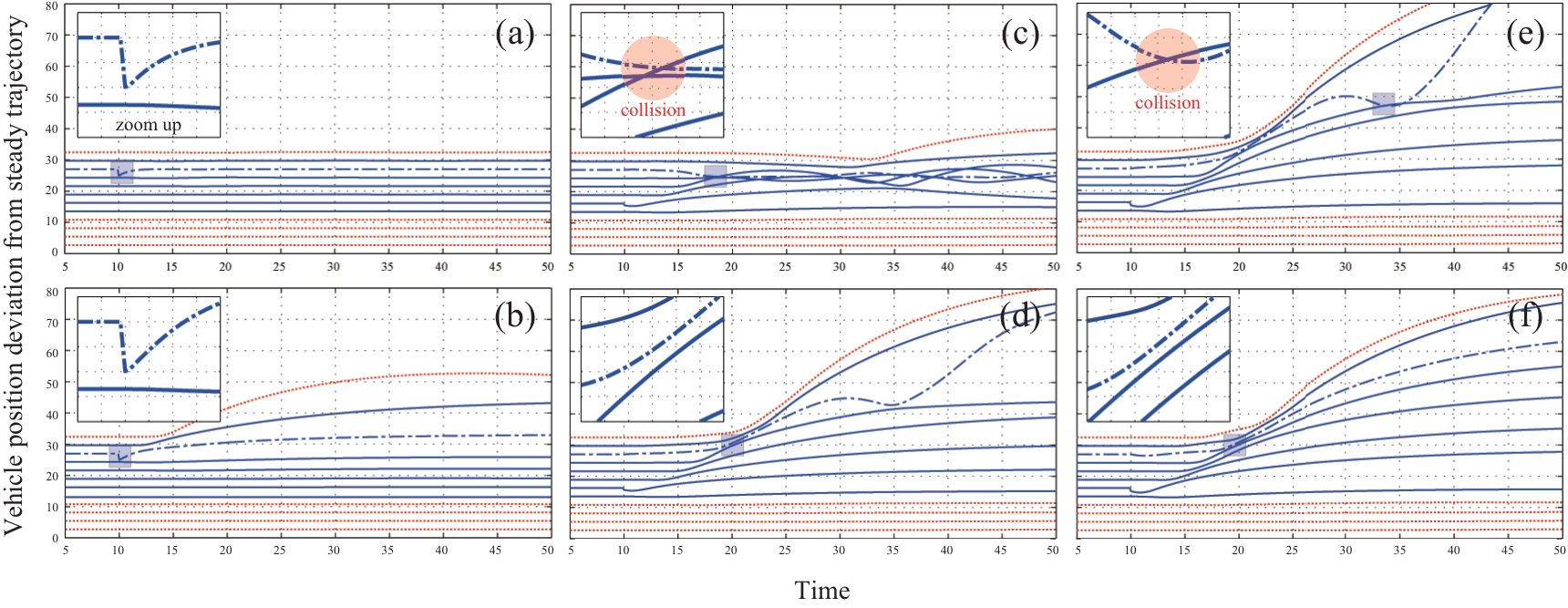}
\end{center}
\vspace{-4pt}
\caption{\scriptsize{
Deviation in vehicle position from a steady trajectory.
A close-up view of the shadowed area is provided in each subfigure.
Subfigures (a) and (b) show, respectively, the resultant system responses when we implement the state feedback controller without the output rectifier and the 12-dimensional retrofit controller, where the velocity of Vehicle 10 becomes zero at $t=10$.
Subfigures (c) and (d) correspond to the cases where the same controllers as those in (a) and (b) are used and the velocity of Vehicle 6 decreases by 30\%  at $t=10$.
Subfigures (e) and (f) correspond to the cases where the 12-dimensional and 4-dimensional retrofit controllers are used and the velocity of Vehicle 6 decreases by 60\%.}
}
\label{figallplots}
\end{figure*}

Figs.~\ref{figallplots}~(a) and (b) show, respectively, the resultant system responses when we implement the state feedback controller without the output rectifier, namely
$u_{1}=\mbox{\boldmath $\hat{K}$}_{1}\mbox{\boldmath $P$}_{1}^{\dagger}x_{1}$,
and the 12-dimensional retrofit controller in (\ref{retex}).
Both subfigures show the deviation from the steady trajectory, i.e., $p_{i}(t)-\overline{v}t$, when the velocity of Vehicle~10 becomes zero at time $t=10$ due to sudden braking.
The blue chained line corresponds to Vehicle~10, the blue solid lines correspond to Vehicles~5--9 and 11, and the red dotted lines correspond to the other vehicles.
From these figures, we can see that both controllers work well in terms of collision avoidance.
However, as shown in Figs.~\ref{figallplots}~(c) and (d), where the velocity of Vehicle~6 is supposed to decrease by 30\%, the feedback controller without the output rectifier induces a collision whereas the retrofit controller does not.
This is because the retrofit controller retains the stability of the original system involving the favorable nonlinearity of $f$ and $g$ in (\ref{dynpltn}), which prevents collision accidents owing to driver operation.
Note that the positions of vehicles in Fig.~\ref{figallplots}~(d) gradually return to their steady trajectories.

Next, we consider reducing the dimension of the retrofit controller from 12.
In the following, $\mbox{\boldmath $P$}_{1}$ and $\mbox{\boldmath $P$}_{1}^{\dagger}$ are determined based on balanced truncation \citep{antoulas2005approximation}, which is used to extract a dominant controllable subspace of $\Sigma_{1}$.
Assuming that the velocity of Vehicle~6 decreases by 60\%, which exceeds the scenario in Fig.~\ref{figallplots}~(d), the resultant system responses in Figs.~\ref{figallplots}~(e) and (f) correspond to the 12-dimensional and 4-dimensional retrofit controllers, respectively.
From these figures, we see that the 4-dimensional retrofit controller can avoid a collision but the 12-dimensional controller cannot.

The reason for this outcome can be explained as follows.
The 12-dimensional retrofit controller is forced to use state feedback information from Vehicles~5--11, irrespective of the distance of these vehicles from the 10th controlled vehicle.
Because vehicles that are distant from the input port are not sufficiently controllable, feedback control based on the measurement of such weakly controllable states may induce oscillatory behavior in a closed-loop system.
Conversely, the low-dimensional controller can naturally focus its attention on the dominant controllable subspace.
This is because, through model reduction, we can eliminate the subspace that is approximately uncontrollable.
Thus, the model reduction technique can be regarded as a systematic tool to extract such a dominant controllable subspace.
This example highlights that low-dimensional retrofit controllers, as opposed to higher-dimensional ones, are more reasonable when the number of actuators is limited.

\section{Concluding Remarks}\label{seccr}

In this paper, we have proposed a retrofit control method for stable linear and nonlinear network systems.
The proposed method only requires a model of the subsystem of interest for controller design.
The resultant retrofit controller is implemented as a cascade interconnection of a local controller that stabilizes an isolated model of the subsystem of interest and a dynamical output rectifier that rectifies an output signal of the subsystem so as to conform to an output signal of the isolated subsystem model while acquiring complementary signals neglected in local controller design, such as interconnection and nonlinear feedback signals.

Future work will consider the generalization of the proposed scheme to robust control under consideration of modeling error in the local subsystem.
Another important future work is to devise a method to determine a reasonable set of subsystems for a given network system.
Indeed, the resultant control performance should be dependent on several factors: for example, subsystem partition, the number of subsystems, and the location of retrofit controllers to be implemented.
Even though the determination of them may require some global knowledge of network systems,  utilizing such a global system knowledge for local retrofit controller design would be beneficial to attain better control performance.

\bibliographystyle{abbrvnat}         
\bibliography{IEEEabrv,reference,reference_CREST}            

\begin{thebibliography}{34}
\providecommand{\natexlab}[1]{#1}
\providecommand{\url}[1]{\texttt{#1}}
\expandafter\ifx\csname urlstyle\endcsname\relax
  \providecommand{\doi}[1]{doi: #1}\else
  \providecommand{\doi}{doi: \begingroup \urlstyle{rm}\Url}\fi

\bibitem[Antoulas(2005)]{antoulas2005approximation}
A.~C. Antoulas.
\newblock \emph{Approximation of large-scale dynamical systems}.
\newblock Society for Industrial and Applied Mathematics, Philadelphia, PA,
  USA, 2005.
\newblock ISBN 0898715296.

\bibitem[Bamieh et~al.(2002)Bamieh, Paganini, and
  Dahleh]{bamieh2002distributed}
B.~Bamieh, F.~Paganini, and M.~A. Dahleh.
\newblock Distributed control of spatially invariant systems.
\newblock \emph{Automatic Control, IEEE Transactions on}, 47\penalty0
  (7):\penalty0 1091--1107, 2002.

\bibitem[Bernstein(2009)]{bernstein2009matrix}
D.~S. Bernstein.
\newblock \emph{Matrix mathematics: theory, facts, and formulas}.
\newblock Princeton University Press, 2009.

\bibitem[Blondel and Tsitsiklis(2000)]{blondel2000survey}
V.~D. Blondel and J.~N. Tsitsiklis.
\newblock A survey of computational complexity results in systems and control.
\newblock \emph{Automatica}, 36\penalty0 (9):\penalty0 1249--1274, 2000.

\bibitem[Chakrabortty(2011)]{chakrabortty2011wide}
A.~Chakrabortty.
\newblock Wide-area damping control of large power systems using a model
  reference approach.
\newblock In \emph{Decision and Control, held jointly with European Control
  Conference (CDC-ECC), 2011 Proceedings of the 50th IEEE Conference on}, pages
  2189--2194. IEEE, 2011.

\bibitem[D'Andrea and Dullerud(2003)]{d2003distributed}
R.~D'Andrea and G.~E. Dullerud.
\newblock Distributed control design for spatially interconnected systems.
\newblock \emph{Automatic Control, IEEE Transactions on}, 48\penalty0
  (9):\penalty0 1478--1495, 2003.

\bibitem[Ebihara et~al.(2012)Ebihara, Peaucelle, and
  Arzelier]{ebihara2012decentralized}
Y.~Ebihara, D.~Peaucelle, and D.~Arzelier.
\newblock Decentralized control of interconnected positive systems using
  {$L_1$}-induced norm characterization.
\newblock In \emph{Decision and Control (CDC), 2012 Proceedings of the 51st
  IEEE Conference on}, pages 6653--6658. IEEE, 2012.

\bibitem[Farokhi and Johansson(2015)]{farokhi2015optimal}
F.~Farokhi and K.~H. Johansson.
\newblock Optimal control design under limited model information for
  discrete-time linear systems with stochastically-varying parameters.
\newblock \emph{IEEE Transactions on Automatic Control}, 60\penalty0
  (3):\penalty0 684--699, 2015.

\bibitem[Farokhi et~al.(2013)Farokhi, Langbort, and
  Johansson]{farokhi2013optimal}
F.~Farokhi, C.~Langbort, and K.~H. Johansson.
\newblock Optimal structured static state-feedback control design with limited
  model information for fully-actuated systems.
\newblock \emph{Automatica}, 49\penalty0 (2):\penalty0 326--337, 2013.

\bibitem[Hayakawa and Nakanishi(1998)]{hayakawa1998theory}
H.~Hayakawa and K.~Nakanishi.
\newblock Theory of traffic jam in a one-lane model.
\newblock \emph{Physical Review E}, 57\penalty0 (4):\penalty0 3839, 1998.

\bibitem[Hill and Moylan(1978)]{moylan1978stability}
D.~Hill and P.~Moylan.
\newblock Stability criteria for large-scale systems.
\newblock \emph{Automatic Control, IEEE Transactions on}, 23\penalty0
  (2):\penalty0 143--149, 1978.

\bibitem[{\.I}ftar(1993)]{iftar1993decentralized}
A.~{\.I}ftar.
\newblock Decentralized estimation and control with overlapping input, state,
  and output decomposition.
\newblock \emph{Automatica}, 29\penalty0 (2):\penalty0 511--516, 1993.

\bibitem[Ikeda et~al.(1984)Ikeda, {\v{S}}iljak, and White]{ikeda1984inclusion}
M.~Ikeda, D.~D. {\v{S}}iljak, and D.~E. White.
\newblock An inclusion principle for dynamic systems.
\newblock \emph{Automatic Control, IEEE Transactions on}, 29\penalty0
  (3):\penalty0 244--249, 1984.

\bibitem[Ilic and Liu(1996)]{ilic1996hierarchical}
M.~D. Ilic and S.~Liu.
\newblock \emph{Hierarchical power systems control: its value in a changing
  industry}.
\newblock Springer Heidelberg, 1996.

\bibitem[Khalil and Grizzle(1996)]{khalil1996nonlinear}
H.~K. Khalil and J.~Grizzle.
\newblock \emph{Nonlinear systems}, volume~3.
\newblock Prentice hall New Jersey, 1996.

\bibitem[Kundur(1994)]{kundur1994power}
P.~Kundur.
\newblock \emph{Power system stability and control}.
\newblock Tata McGraw-Hill Education, 1994.

\bibitem[Langbort and Delvenne(2010)]{langbort2010distributed}
C.~Langbort and J.~Delvenne.
\newblock Distributed design methods for linear quadratic control and their
  limitations.
\newblock \emph{Automatic Control, IEEE Transactions on}, 55\penalty0
  (9):\penalty0 2085--2093, 2010.

\bibitem[Langbort et~al.(2004)Langbort, Chandra, and
  D'Andrea]{langbort2004distributed}
C.~Langbort, R.~S. Chandra, and R.~D'Andrea.
\newblock Distributed control design for systems interconnected over an
  arbitrary graph.
\newblock \emph{Automatic Control, IEEE Transactions on}, 49\penalty0
  (9):\penalty0 1502--1519, 2004.

\bibitem[Qu and Simaan(2014)]{qu2014modularized}
Z.~Qu and M.~A. Simaan.
\newblock Modularized design for cooperative control and plug-and-play
  operation of networked heterogeneous systems.
\newblock \emph{Automatica}, 50\penalty0 (9):\penalty0 2405--2414, 2014.

\bibitem[Rantzer(2015)]{rantzer2015scalable}
A.~Rantzer.
\newblock Scalable control of positive systems.
\newblock \emph{European Journal of Control}, 24:\penalty0 72--80, 2015.

\bibitem[Rotkowitz and Lall(2006)]{rotkowitz2006characterization}
M.~Rotkowitz and S.~Lall.
\newblock A characterization of convex problems in decentralized control.
\newblock \emph{Automatic Control, IEEE Transactions on}, 51\penalty0
  (2):\penalty0 274--286, 2006.

\bibitem[Sadamoto et~al.(2014)Sadamoto, Ishizaki, and
  Imura]{sadamoto2014hierarchical}
T.~Sadamoto, T.~Ishizaki, and J.-i. Imura.
\newblock Hierarchical distributed control for networked linear systems.
\newblock In \emph{Decision and Control (CDC), 2014 IEEE 53rd Annual Conference
  on}, pages 2447--2452. IEEE, 2014.

\bibitem[Sadamoto et~al.(2016)Sadamoto, Ishizaki, Imura, Sandberg, and
  Johansson]{sadamoto2016retrofitting}
T.~Sadamoto, T.~Ishizaki, J.-i. Imura, H.~Sandberg, and K.~H. Johansson.
\newblock Retrofitting state feedback control of networked nonlinear systems
  based on hierarchical expansion.
\newblock In \emph{Decision and Control (CDC), 2016 IEEE 55th Annual Conference
  on}, pages 3432--3437. IEEE, 2016.

\bibitem[Sepulchre et~al.(2012)Sepulchre, Jankovic, and
  Kokotovic]{sepulchre2012constructive}
R.~Sepulchre, M.~Jankovic, and P.~V. Kokotovic.
\newblock \emph{Constructive nonlinear control}.
\newblock Springer Science \& Business Media, 2012.

\bibitem[{\v{S}}iljak(1972)]{siljak1972stability}
D.~D. {\v{S}}iljak.
\newblock Stability of large-scale systems under structural perturbations.
\newblock \emph{Systems, Man, and Cybernetics, IEEE Transactions on},
  SMC-2\penalty0 (5):\penalty0 657--663, 1972.

\bibitem[{\v{S}}iljak(1991)]{siljak1991decentralized}
D.~D. {\v{S}}iljak.
\newblock \emph{Decentralized control of complex systems}, volume 184.
\newblock Mathematics in Science and Engineering, Academic Press, 1991.

\bibitem[{\v{S}}iljak and Ze{\v{c}}evi{\'c}(2005)]{vsiljak2005control}
D.~D. {\v{S}}iljak and A.~I. Ze{\v{c}}evi{\'c}.
\newblock Control of large-scale systems: Beyond decentralized feedback.
\newblock \emph{Annual Reviews in Control}, 29\penalty0 (2):\penalty0 169--179,
  2005.

\bibitem[Stipanovi{\'c} et~al.(2004)Stipanovi{\'c}, Inalhan, Teo, and
  Tomlin]{stipanovic2004decentralized}
D.~M. Stipanovi{\'c}, G.~Inalhan, R.~Teo, and C.~J. Tomlin.
\newblock Decentralized overlapping control of a formation of unmanned aerial
  vehicles.
\newblock \emph{Automatica}, 40\penalty0 (8):\penalty0 1285--1296, 2004.

\bibitem[Tan and Ikeda(1990)]{tan1990decentralized}
X.-L. Tan and M.~Ikeda.
\newblock Decentralized stabilization for expanding construction of large-scale
  systems.
\newblock \emph{IEEE Transactions on automatic control}, 35\penalty0
  (6):\penalty0 644--651, 1990.

\bibitem[Wang and Davison(1973)]{wang1973stabilization}
S.-H. Wang and E.~Davison.
\newblock On the stabilization of decentralized control systems.
\newblock \emph{IEEE Transactions on Automatic Control}, 18\penalty0
  (5):\penalty0 473--478, 1973.

\bibitem[Wang et~al.(1995)Wang, Xie, and de~Souza]{wang1995robust}
Y.~Wang, L.~Xie, and C.~E. de~Souza.
\newblock Robust decentralized control of interconnected uncertain linear
  systems.
\newblock In \emph{Decision and Control, 1995., Proceedings of the 34th IEEE
  Conference on}, volume~3, pages 2653--2658. IEEE, 1995.

\bibitem[Willems(1972{\natexlab{a}})]{willems1972dissipativeI}
J.~C. Willems.
\newblock Dissipative dynamical systems part {I}: General theory.
\newblock \emph{Archive for Rational Mechanics and Analysis}, 45\penalty0
  (5):\penalty0 321--351, 1972{\natexlab{a}}.

\bibitem[Willems(1972{\natexlab{b}})]{willems1972dissipativeII}
J.~C. Willems.
\newblock Dissipative dynamical systems part {II}: Linear systems with
  quadratic supply rates.
\newblock \emph{Archive for Rational Mechanics and Analysis}, 45\penalty0
  (5):\penalty0 352--393, 1972{\natexlab{b}}.

\bibitem[Zhou et~al.(1996)Zhou, Doyle, and Glover]{zhou1996robust}
K.~Zhou, J.~C. Doyle, and K.~Glover.
\newblock \emph{Robust and optimal control}, volume~40.
\newblock Prentice Hall New Jersey, 1996.

\end{thebibliography}



\end{document}